\documentclass[twocolumn,twocolappendix,tighten,trackchanges]{aastex7}

\usepackage{booktabs}
\usepackage{float}
\usepackage{multirow}
\usepackage{diagbox}
\usepackage[english]{babel}

\addto\extrasenglish{%
}

\newcommand{\ifm}[1]{\relax\ifmmode#1\else$\mathsurround=0pt #1$\fi}

\def\dex{{\rm\thinspace dex}}

\def\m{{\rm\thinspace m}}

\def\pc{{\rm\thinspace pc}}
\def\kpc{{\rm\thinspace kpc}}

\def\Mpc{{\rm\thinspace Mpc}}
\def\Gpc{{\rm\thinspace Gpc}}

\newcommand{\hMpc}{\,\ifm{h^{-1}}{\rm Mpc}}
\newcommand{\hkpc}{\,\ifm{h^{-1}}{\rm kpc}}
\def\kms{{\rm\thinspace km\thinspace s}^{-1}}

\def\pcmcb{{\rm\thinspace cm}^{-3}}

\def\g{{\rm\thinspace g}}

\def\Msun{\hbox{$\rm\thinspace M_{\odot}$}}
\newcommand{\hMsun}{\,\ifm{h^{-1}}{\rm\thinspace M_{\odot}}}

\def\s{{\rm\thinspace s}}       
\def\yr{{\rm\thinspace yr}}

\def\Msunpc2{{\Msun\pc}^{-2}}
\def\Msunyrkpc2{{\Msun\yr^{-1}\kpc}^{-2}}

\def\magarcsec2{{\rm\thinspace mag\thinspace arcsec}^{-2}}

\shorttitle{2nd-Generation CAMELS: $(50\hMpc)^3$ boxes}
\shortauthors{Genel et al.}
\graphicspath{{./}{figures/}}

\begin{document}

\title{Learning the Universe with the 2$^{\rm nd}$ Generation of CAMELS:\\ Varying 35 parameters of the IllustrisTNG model in $(50\hMpc)^3$ boxes}

\author[0000-0002-3185-1540]{Shy Genel}
\affiliation{Center for Computational Astrophysics, Flatiron Institute, 162 5th Avenue, New York, NY 10010, USA}
\affiliation{Columbia Astrophysics Laboratory, Columbia University, 550 West 120th Street, New York, NY 10027, USA}
\email[show]{sgenel@flatironinstitute.org}
\author[0000-0003-3977-1761]{Yongseok Jo}
\affiliation{NSF-Simons AI Institute for the Sky (SkAI), 172 E. Chestnut St., Chicago, IL 60611, USA}
\email[]{me@yongseokjo.com}
\author[orcid=0000-0003-4597-6739]{Boon Kiat Oh}
\affiliation{School of Physics, Korea Institute for Advanced Study, 85 Hoegiro, Dongdaemun-gu, Seoul 02455, Republic of Korea}
\affiliation{Department of Physics, University of Connecticut, 196 Auditorium Road, U-3046, Storrs, CT 06269-3046, USA}
\email[]{bkoh@kias.re.kr}
\author[0000-0002-1185-4111]{Megan Taylor Tillman}
\affiliation{Department of Physics and Astronomy, Rutgers University,  136 Frelinghuysen Rd, Piscataway, NJ 08854, USA}
\email[]{mtt74@physics.rutgers.edu}
\author[0000-0002-2318-3087]{Max E. Lee}
\affiliation{Department of Astronomy, Columbia University, MC 5246, 538 West 120th Street, New York, NY 10027, USA}
\email[]{max.e.lee@columbia.edu}
\author[0009-0006-4981-0604]{Jun-Young Lee}
\affiliation{Department of Astrophysical Sciences, Princeton University, 4 Ivy Lane, Princeton, NJ 08544, USA}
\email[]{junyoung.lee@princeton.edu}
\author[0000-0002-1329-9246]{Elena Hern\'andez-Mart\'inez}
\affiliation{TransferLab, appliedAI Institute for Europe, Freddie-Mercury-Straße 5, 80797 München, Germany}
\email[]{e.hernandez@appliedai-institute.de}
\author[0000-0001-7964-5933]{Christopher C. Lovell}
\affiliation{Kavli Institute for Cosmology, Madingley Road, Cambridge CB3 0HA, UK}
\affiliation{Institute of Astronomy, Madingley Road, Cambridge CB3 0HA, UK}
\email[]{chris.lovell.astro@gmail.com}
\author[0009-0000-6189-099X]{Xavier Sims}
\affiliation{Department of Physics, University of Connecticut, 196 Auditorium Road, U-3046, Storrs, CT 06269-3046, USA}
\email[]{xavier.sims@uconn.edu}
\author[0000-0001-5817-5944]{Blakesley Burkhart}
\affiliation{Center for Computational Astrophysics, Flatiron Institute, 162 5th Avenue, New York, NY 10010, USA}
\affiliation{Department of Physics and Astronomy, Rutgers University,  136 Frelinghuysen Rd, Piscataway, NJ 08854, USA}
\email[]{bburkhart@flatironinstitute.org}
\author[0000-0001-7457-8487]{Kentaro Nagamine}
\affiliation{Theoretical Astrophysics, Department of Earth and Space Science, Graduate School of Science, The University of Osaka, 1-1 Machikaneyama, Toyonaka, Osaka, 560-0043, Japan}
\affiliation{Theoretical Joint Research, Forefront Research Center, The University of Osaka, Toyonaka, Osaka 560-0043, Japan}
\affiliation{Kavli IPMU (WPI), UTIAS, The University of Tokyo, Kashiwa, Chiba 277-8583, Japan}
\affiliation{Department of Physics and Astronomy, University of Nevada, Las Vegas, 4505 S. Maryland Pkwy, Las Vegas, NV 89154-4002, USA}
\affiliation{Nevada Center for Astrophysics, University of Nevada, Las Vegas, 4505 S. Maryland Pkwy, Las Vegas, NV 89154-4002, USA}
\email[]{kn@astro-osaka.jp}
\author[0000-0001-5769-4945]{Daniel Angl\'es-Alc\'azar}
\affiliation{Department of Physics, University of Connecticut, 196 Auditorium Road, U-3046, Storrs, CT 06269-3046, USA}
\email[]{angles-alcazar@uconn.edu}
\author[0000-0002-4816-0455]{Francisco Villaescusa-Navarro}
\affiliation{Center for Computational Astrophysics, Flatiron Institute, 162 5th Avenue, New York, NY 10010, USA}
\email[]{fvillaescusa@flatironinstitute.org}

\begin{abstract}
We present a new set of 1,192 cosmological simulations as part of the CAMELS project, in which a space of 35 cosmological, astrophysical, and numerical parameters is explored around the fiducial IllustrisTNG model.
The volume of each of these simulations is $(50\hMpc)^3$, eight times larger than that of previous CAMELS simulations. This provides lower sample variance as well as access to more massive halos and more diverse environments. We focus this work on exploring the advantages these differences provide for parameter inference powered by neural networks.
We generate training sets based on the matter power spectra, projected maps of the volumes, graphs representing galaxy spatial distributions, and thermodynamical properties of massive halos.
We employ multilayer perceptrons, convolutional neural networks, graph neural networks, and Gaussian processes, respectively, to extract information on the simulation parameters from these inputs while comparing systematically to analogous results from our previous generation of $(25\hMpc)^3$ simulations.
We generally find that the new, larger volumes produce tighter marginal constraints on the parameters, to degrees that vary between the different inputs. The improvements, however, scale more weakly than with the square root of the increase in the amount of data (i.e., physical volume). We interpret this as originating either from information loss due to mode coupling or from complex degeneracies in parameter space.
We also discuss the effects on statistics of the intergalactic medium temperature from four new parameters that are varied in these simulations, which control the amplitude and timing of the ionizing background radiation.
We publicly release the simulation outputs and ancillary data at \url{https://camels.readthedocs.io}.
\end{abstract}


\section{Introduction}
\label{sec:introduction}

The standard model of cosmology, the $\Lambda$CDM paradigm, has been remarkably successful in explaining a wide range of observations, from the cosmic microwave background (CMB) to the large-scale distribution of galaxies \citep{SpringelV_06a,HinshawG_13a,AndersonL_14a,Planck2018}. In this model, a handful of parameters characterize the Universe’s composition and initial conditions, such as the present-day matter density fraction $\Omega_m$ and the amplitude of matter fluctuations on $8\hMpc$ scales $\sigma_8$. With robust theoretical predictions at hand, observational data, such as that from CMB experiments and galaxy surveys, have allowed for the measurement of these with a few-percent accuracy \citep{Planck2018,AlamS_21a,AbbottT_22a}, utilizing information on cosmic scales whose evolution is at most mildly non-linear and not significantly affected by galaxy formation physics. Continued improvements in observational precision, for instance, with missions like LSST and \emph{Euclid} \citep{LSSTScienceBook,Euclid2025}, demand equally robust theoretical predictions to extract further information from smaller scales.
Tightening the constraints on cosmological parameters and initial conditions has the potential to allow detections of subtle deviations from $\Lambda$CDM that would point towards new physics.

Numerical simulations have become indispensable tools for modeling the non-linear formation of structure in the Universe. While large-volume $N$-body simulations have mapped out the cosmic web of dark matter on enormous scales \citep{SpringelV_05a,KlypinA_10a}, purely gravitational calculations cannot capture the complex baryonic processes that shape the Universe. In the past decade, hydrodynamical simulations have incorporated gas dynamics, radiative cooling, star formation, and feedback processes into large cosmological volumes, revolutionizing our understanding of galaxy formation within the large-scale structure. Projects like IllustrisTNG \citep{MarinacciF_17a,NaimanJ_17a,NelsonD_17a,PillepichA_17a,SpringelV_17a}, EAGLE \citep{SchayeJ_14a,CrainR_15a}, and SIMBA \citep{DaveR_19a} have achieved $\sim100\Mpc$-scale volumes with baryonic physics calibrated to reproduce key observables such as galaxy stellar masses and star formation rates. These simulations have elucidated the impact of energetic feedback from supernovae and active galactic nuclei (AGN) in regulating galaxy growth and altering the surrounding gas, thereby modulating the matter distribution on small to intermediate scales \citep{vanDaalenM_20a}.

More recently, even larger hydrodynamical simulations are being run to narrow the remaining gap with observational survey volumes, enabling direct predictions for upcoming cluster counts and lensing surveys \citep{PakmorR_23a,SchayeJ_23a,KugelR_25a}. Despite their successes, a common limitation of such flagship simulations is that they typically assume a single fiducial cosmology and a fixed or limited set of subgrid physics parameters. This makes it challenging to assess the effects of varying fundamental parameters or to quantify uncertainties due to galaxy formation physics, as running additional large simulations for each parameter choice is computationally prohibitive.

While simulation projects targeting the exploration of physics variations have been run before \citep{SchayeJ_10a,TorreyP_14a}, the advent of machine learning has enabled a paradigm shift embodied in the CAMELS project (\textit{Cosmology and Astrophysics with MachinE Learning Simulations}). CAMELS was conceived to explore cosmological and astrophysical parameter spaces more efficiently \citep{Villaescusa-NavarroF_21a} with a novel approach: instead of one huge simulation or dozens of simulations in which certain parameters are varied, it comprises thousands of medium-resolution, smaller-volume simulations designed specifically for machine learning and parameter inference applications. The primary goal of CAMELS has been to enable machine learning algorithms to learn the complex mappings between observables (simulated galaxy catalogs, 2D or 3D maps of matter or neutral hydrogen fields, etc.) and the underlying input parameters. By sampling a broad range of these parameters, CAMELS created a large training set of diverse universes that allows for the marginalization of baryonic effects for cosmological parameter inference, particularly when using small scales that are heavily affected by baryons and cannot be treated analytically. Early studies using the CAMELS dataset demonstrated the power of ML techniques -- for example, showing that convolutional neural networks could predict $\Omega_m$ from simulated 2D density maps far more accurately than traditional summary statistics \citep{Villaescusa-NavarroF_22c,JoY_25a}. These promising results illustrate how an abundance of simulated realizations can help train models to recognize subtle features in the data that correlate with different cosmological or astrophysical parameters.

Similarly, the ANTILLES suite, which consists of 400 cosmological hydrodynamical simulations that vary stellar and AGN feedback efficiencies as well as hydrodynamics formulations over a wide range, enables empirical modeling of how feedback impacts the nonlinear matter power spectrum \citep{SalcidoJ_23a}. This approach demonstrates how a moderately sized ensemble, designed with parameter variation in mind, can yield informative bounds on baryonic effects approaching percent-level precision. The FLAMINGO suite similarly includes variations of subgrid model parameters with the goal of spanning a large range of possible physics implementations in very large, $\Gpc$-scale cosmological volumes \citep{SchayeJ_23a}. The sheer size of each of these simulations does not allow for a large number of them that would be amenable to training machine learning models; however, their calibration used Gaussian processes trained on smaller volumes that sampled the parameter space more densely \citep{KugelR_23a}.

The original CAMELS suite focused on varying two cosmological parameters ($\Omega_m$ and $\sigma_8$) along with a handful of galaxy formation parameters that control the efficiency of feedback from supernovae and AGN. It comprised simulations run using two state-of-the-art hydrodynamic codes that employed the IllustrisTNG and SIMBA galaxy formation models \citep{Villaescusa-NavarroF_21a}. This dual-code strategy was adopted to marginalize over uncertain aspects of subgrid physics, as any robust inference method should ideally perform well across different implementations of baryonic physics. Later, the project was extended to include a third suite, CAMELS-Astrid \citep{NiY_23a}, which employs the galaxy formation model of the Astrid simulation \citep{BirdS_22a,NiY_22a} and contains over 2,100 simulations varying three cosmological parameters ($\Omega_m$, $\sigma_8$, and $\Omega_b$) and four feedback parameters. This broader suite enables improved extrapolation performance and greater robustness of machine learning models by sampling a wider diversity of galaxy formation behaviors. \citet{NiY_23a} also extended the parameter space to include a much larger and more complete parameter space of the IllustrisTNG model, varying 22 additional parameters on top of the original 6 that were varied in \citet{Villaescusa-NavarroF_21a}, including 3 additional cosmological parameters.

\begin{figure*}[ht!]
\gridline{
  \fig{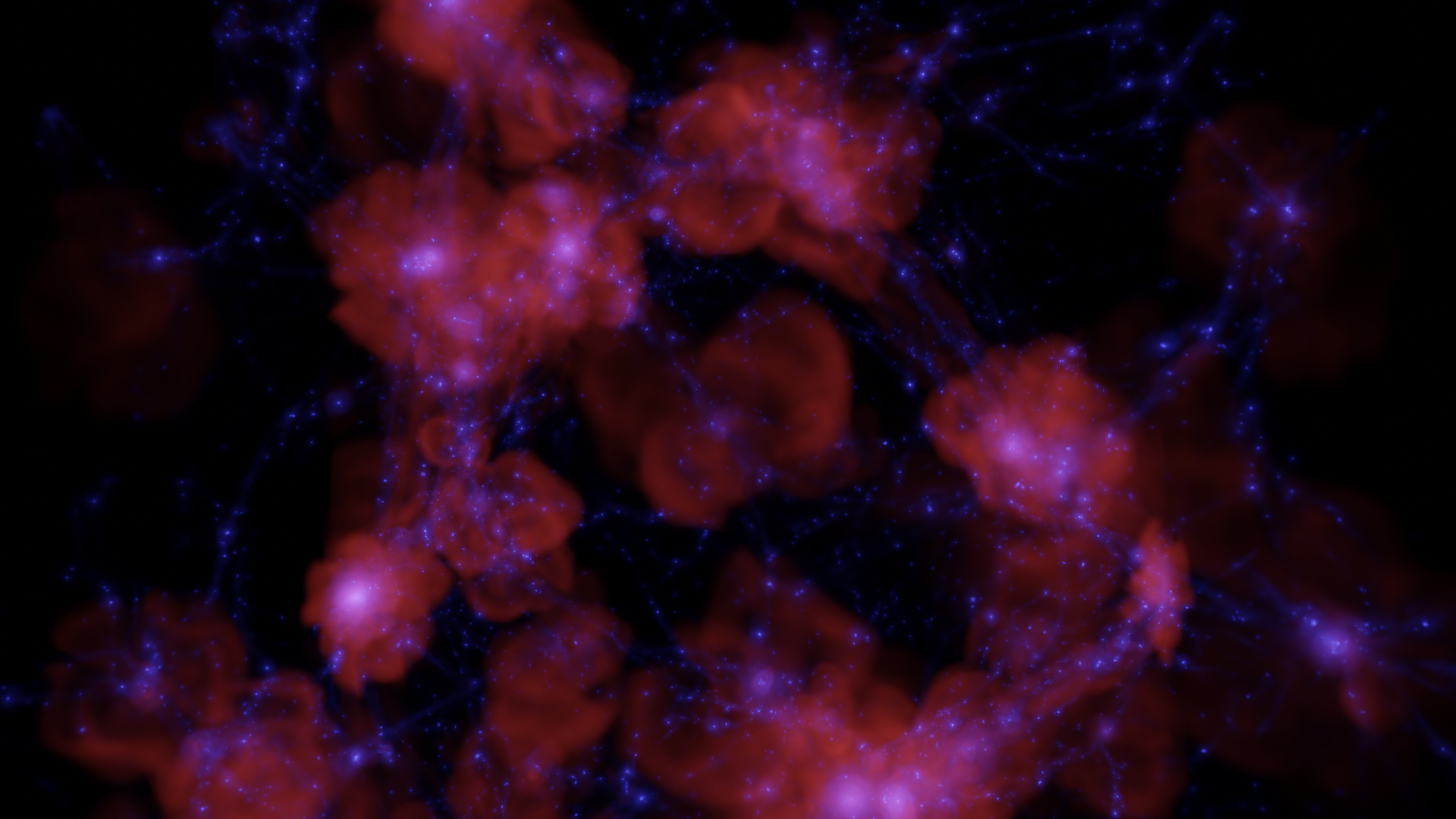}{0.3\textwidth}{(a) $(25\hMpc)^3$}
  \fig{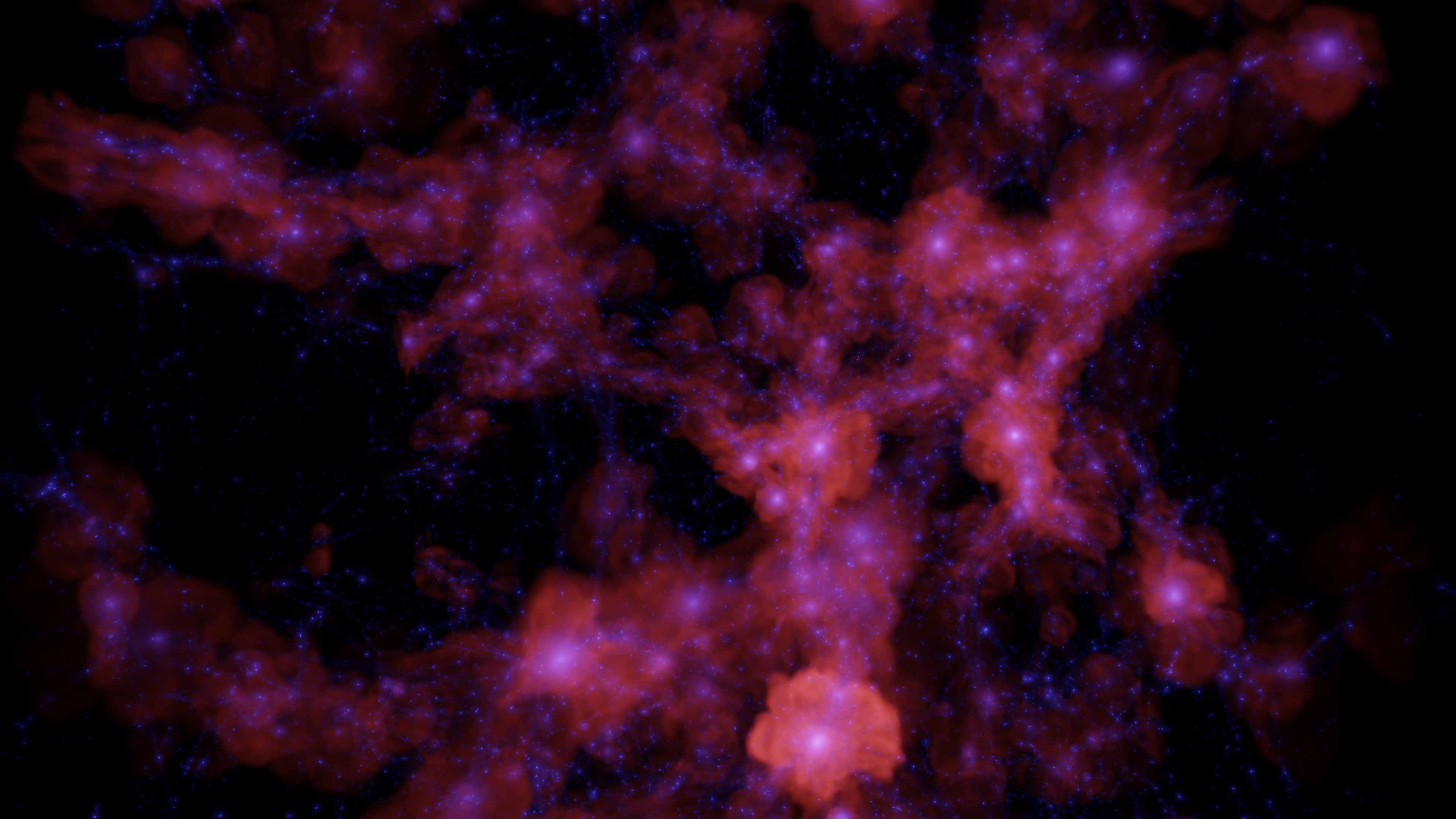}{0.6\textwidth}{(b) $(50\hMpc)^3$}
}
\caption{Density-temperature visualizations of two CAMELS simulations, from the first generation (a) and from the second generation (b), which is presented in this work.}
\label{fig:visualization}
\end{figure*}

While these ``first-generation" CAMELS simulations proved the viability of this approach, they also highlighted several limitations that motivate a second-generation suite of larger simulations. Each CAMELS simulation published so far (besides zoom-in and pure N-body simulations) covers a comoving volume of $(25\hMpc)^3$, which is modest by survey standards \citep{Villaescusa-NavarroF_21a}. A $25\hMpc$ box captures nonlinear structures like galactic halos and groups, but it does not encompass rare, large-scale objects and the full diversity of environments, such as massive galaxy clusters or large voids \citep{KlypinA_10a,SchayeJ_23a}. Additionally, such boxes miss long-wavelength modes of the density field, which generally leads to biased results, e.g.~in the evolution of the power spectrum. The first-generation CAMELS runs also suffer from significant sample variance \citep{NicolaA_22a}. Targeted extensions to CAMELS confronted these challenges by simulating larger volumes \citep{PerezL_23a} or massive halos \citep{LeeM_24a}, but at the cost of, respectively, using semi-analytical models instead of hydrodynamical simulations or zooming in on individual halos instead of simulating full cosmological boxes, which either complicates or limits their utility.

In this paper, we introduce a new second-generation CAMELS simulation suite designed to ameliorate these issues and further advance small-scale studies of structure formation. The new simulations significantly increase the simulated volume to $(50\hMpc)^3$, as illustrated in \autoref{fig:visualization}, while preserving the resolution of the original runs, meaning that the particle number scales up accordingly, and so does the computational cost. In addition to the 5 cosmological parameters and 23 baryonic modeling parameters used in the previously most expansive CAMELS parameter space (the SB28 set;\citealp{NiY_23a}), the new suite introduces several additional astrophysical parameters, with a particular emphasis on those related to the ionizing background radiation. We now explicitly vary the amplitude and timing of the cosmic UV/X-ray ionizing background, which were fixed in previous runs. By doing so, we can assess how uncertainties in the reionization process and photo-heating feedback propagate into observables like the intergalactic medium (IGM) thermal state, the neutral hydrogen distribution, and galaxy properties. The new dataset, therefore, provides an extensive and diverse foundation for studying how changes in fundamental physics and astrophysical processes imprint on nonlinear structure formation.

The scientific aims of this second-generation CAMELS project are broad. At its core, the dataset is meant to serve the community as a public resource for developing and testing algorithms that connect observations to simulation parameters in the presence of complex astrophysical effects.
This work is also part of the ``Learning the Universe" collaboration\footnote{\url{https://learning-the-universe.org}}, whose goal is to infer the fundamental parameters and the initial conditions of the Universe via a statistical inference pipeline that relies on forward models of the distribution and properties of galaxies and cosmic gas. To perform such inference, it is necessary to employ a range of baryonic physics implementations in diverse cosmic environments, which the second-generation CAMELS simulations enable in a more complete way than previously possible.

In this work, we present, using a range of data products from the simulations and correspondingly a range of techniques, an introductory exploration of the constraining power of these larger volumes on cosmological and subgrid physics parameters. In particular, we contrast the results with those that are attainable from our previous, smaller simulation volumes and draw conclusions from these apples-to-apples comparisons. The datasets and techniques presented here have broad relevance, as they pave the way for simulation-based inference in cosmology, whereby entire simulated universes (rather than simplified models) are directly used to train flexible models (like neural networks or Gaussian processes) that can then be applied to real data \citep{HahnC_24a}. This paradigm complements more traditional analytic modeling and has the potential to extract information that lies in the complex, high-order patterns of structure.

This paper is organized as follows. In \autoref{sec:simulations} we describe the new CAMELS simulation suite in detail, including the cosmological and astrophysical parameter space, the IllustrisTNG subgrid model, and resolution and volume characteristics. \autoref{sec:variations} presents global properties of the new simulations, including characterizations of the large-scale structure, the IGM, and galaxy properties, in comparison to previous CAMELS simulations. In \autoref{sec:inference}, we present the results of our inference experiments, including the machine learning and inference methods we apply, comparing the performance of simulation sets of different volumes. \autoref{sec:summary} provides a discussion of the implications of these results and a summary, including information about the public availability of the new simulations for wide community use.

\section{Simulations}
\label{sec:simulations}

\subsection{The IllustrisTNG model and the varied parameters}
\label{sec:TNGmodel}

The IllustrisTNG model is a state-of-the-art framework for incorporating baryonic physics models in cosmological simulations  \citep{WeinbergerR_16a,PillepichA_16a}, implemented in the TreePM-moving-mesh code Arepo  \citep{SpringelV_10a}. Built upon its predecessor, the subgrid physics model used in the Illustris simulation \citep{VogelsbergerM_13a}, it incorporates key processes in galaxy formation physics, with free parameters tuned for the fiducial version of the model as used in the IllustrisTNG simulations \citep{MarinacciF_17a,NaimanJ_17a,NelsonD_17a,PillepichA_17a,SpringelV_17a} to approximately match a number of observed galaxy properties. The CAMELS IllustrisTNG simulations vary both cosmological and subgrid (`astrophysical') parameters to explore their impact on various observables. In the simulations presented here, we vary 35 parameters overall, of which 5 are cosmological. This set of parameters includes all 28 parameters that were varied in our previous runs, which are detailed in \citet{NiY_23a}, as well as 7 additional parameters varied here for the first time.
Some of the key physics components of the IllustrisTNG model, along with the associated parameters that we vary, which are summarized in \autoref{tab:parameters}, are:

\begin{itemize}
\item \textbf{Cosmology}: IllustrisTNG uses cosmological initial conditions, namely a Gaussian random field characterized by a primordial power spectrum whose amplitude is controlled by the $A_s$ parameter (which is converted to the late-time $\sigma_8$ parameter) and slope by the $n_s$ parameter. The simulated system expands as a flat $\Lambda$CDM universe that is characterized by its total matter density $\Omega_m$, its baryonic density $\Omega_b$ and its present-day expansion rate defined by the Hubble parameter $h=H_0/(100\kms/\Mpc)$.

\item \textbf{Self-gravity and Magnetohydrodynamics (MHD)}: Arepo solves the coupled equations of gravity and MHD using a TreePM scheme for the former and a moving Voronoi mesh for the latter. The gravitational forces are softened on small scales to avoid two-body interactions between resolution elements as controlled by the softening parameter $\epsilon$, which has not been previously varied in CAMELS simulations. This numerical parameter is varied in the new simulations presented here since it can be viewed as a nuisance parameter to be marginalized over when performing statistical inference of cosmological or subgrid parameters that carry a more physical meaning.

\item \textbf{Radiative processes:} IllustrisTNG assumes a fiducial \citet{FaucherGiguereC_09a} time-dependent background radiation field, which both ionizes and heats the cosmic gas. The amplitude and time evolution of this background radiation is controlled in our new simulations by 4 parameters following \citet{VillasenorB_22a}, where $\beta_{\rm H}$ and $\beta_{\rm He}$ control the amplitude of the photoionization and photoheating rates of (HI+HeI) and HeII, respectively, while $\Delta z_{\rm H}$ and $\Delta z_{\rm He}$ control the respective timing offsets relative to the fiducial model. The background radiation is spatially uniform except in regions with high gas density, where it is attenuated following \citet{RahmatiA_13a}\footnote{Note that the density at which the \citet{RahmatiA_13a} attenuation kicks in is, in itself, a function of the intensity of the radiation field, an effect that we neglect when we vary the 4 new parameters. Since these parameters give rise to radiation fields that differ by $\lesssim1\dex$ from the fiducial one at any $z\lesssim6$ (which is the range in which the \citet{RahmatiA_13a} correction is applied), the typical attenuation density varies by $\lesssim0.6\dex$, which is comparable to the scatter measured in the \citet{RahmatiA_13a} radiative transfer simulations.}.  Gas is also subject to radiative cooling based on an on-the-fly chemical network for hydrogen and helium, and pre-computed photoionization equilibrium cooling tables for metals.

\item \textbf{ISM pressurization and star‐formation}: IllustrisTNG uses an effective equation of state (eEOS) for unresolved multiphase ISM, supporting pressure against collapse. The eEOS is an interpolation controlled by the parameter $q_{\rm eEOS}$, between the \citet{SpringelV_03a} eEOS and an isothermal EOS at $T=10^4K$. Gas above a hydrogen number density threshold of $0.013\pcmcb$ forms stars stochastically following a Schmidt law whose normalization is controlled by the parameter $t_0^*$, the star-formation timescale at the density threshold.

\item \textbf{Stellar evolution}: IllustrisTNG models star particles as simple stellar populations with a Chabrier-like IMF whose high-mass slope ($>1\Msun$) is controlled by the $b_{\rm IMF}$ parameter. The model implements gradual mass return from stars to the gas phase and tracks the production of 9 chemical elements from Type II/Ia SNe and AGB stars. The metal production rate from SNII is modulated by the parameter $M_{\rm SNII,min}$, which determines the minimum stellar mass for stars to explode in a supernova. Two new parameters that we vary in the simulations presented here are the normalization of the SNIa rate $N_0$ as well as the power-law index $s$ of its delay time distribution. 

\item \textbf{Galactic winds}: Stellar feedback drives galactic winds in IllustrisTNG, whose launch speed scales with the local velocity dispersion, with a normalization controlled by the parameter $\kappa_{\rm w}$ and a floor set by $v_{\rm w,min}$. The normalizations of the total energy and momentum fluxes of the winds per stellar mass formed are controlled directly by the parameters $\bar{e}_w$ and $\bar{m}_w$, respectively, as well as indirectly by the parameter $M_{\rm SNII,min}$ mentioned above. The energy flux also scales with local metallicity, following a functional form that includes the three variable parameters $f_{\rm w,Z}$, $Z_{\rm w,ref}$, and $\gamma_{\rm w,Z}$. The winds are hydrodynamically decoupled until the ambient density drops below a value set by the parameter $n_{\rm w,rec}$, and they are largely kinetic but contain a subdominant thermal component whose energy fraction is controlled by the parameter $\tau_w$. A fraction $1-\gamma_{\rm w}$ of the metal content of launched wind particles gets first deposited in the ambient ISM.

\item \textbf{Black Hole (BH) Seeding and Growth}: IllustrisTNG includes a model for the formation and growth of supermassive black holes at the centers of galaxies. Black holes with an initial mass set by the parameter $M_{\rm seed}$ are seeded in halos once they cross a threshold mass of $5\times10^{10}\Msun/h$. Their growth follows a Bondi-like accretion rate with a normalization controlled by the parameter $A_{\rm Bondi}$ that is capped at a multiple $A_{\rm Edd}$ of the Eddington rate.

\item \textbf{Active Galactic Nuclei (AGN) Feedback}: IllustrisTNG employs a dual-mode AGN feedback model, where the choice between the two modes is made based on an Eddington ratio threshold that is a function of the BH mass. The value of that threshold at the `pivot' BH mass of $10^8\Msun$ is set by the parameter $\xi_0$ and its dependence on BH mass is controlled by the parameter $\beta_\chi$. The high accretion mode operates through continuous thermal injections whose energy content is controlled by the radiative efficiency parameter $\epsilon_r$ and by the energy coupling parameter $\epsilon_{\rm f,high}$. The low accretion mode operates through sporadic kinetic injections whose energy content is modulated by the $A_{\rm AGN1}$ and $\epsilon_r$ parameters and burstiness by the parameter $f_{\rm re}$. In parallel to these two modes, a radiative mode adds to the incident radiation field assumed for any gas cell that is located within three times the virial radius of a halo that contains an accreting BH. This additional radiation field modifies the heating and cooling rates of these cells, but in practice its effects on the evolution of the host galaxies and their environments are sub-dominant to the other feedback modes.
\end{itemize}

\begin{table*}[h]
\centering
\begin{tabular}{lccclcl}
\toprule
\multicolumn{1}{l}{Parameter} & \multicolumn{1}{c}{Fiducial} & \multicolumn{1}{c}{Min} & \multicolumn{1}{c}{Max} & \multicolumn{1}{l}{Astrophysical} & \multicolumn{1}{c}{parameter} & \multicolumn{1}{l}{\citet{NiY_23a}} \\
\multicolumn{1}{l}{} & \multicolumn{1}{l}{} & \multicolumn{1}{l}{} & \multicolumn{1}{l}{} & \multicolumn{1}{l}{Context} & \multicolumn{1}{c}{serial number} & \multicolumn{1}{l}{parameter name} \\
\midrule
$\Omega_m$ & 0.3 & 0.1 & 0.5 & Cosmology & 1 & Omega0 \\
$\sigma_8$ & 0.8 & 0.6 & 1 & Cosmology & 2 & sigma8 \\
$\Omega_b$ & 0.049 & 0.029 & 0.069 & Cosmology & 7 & OmegaBaryon \\
$h$ & 0.6711 & 0.4711 & 0.8711 & Cosmology & 8 & HubbleParam \\
$n_s$ & 0.9624 & 0.7624 & 1.1624 & Cosmology & 9 & n\_s \\
$\epsilon$ & 2 & 1 & 4 & Gravity & 35 & \bf{SofteningComovingType01} \\
$\beta_{\rm H}$ & 1 & 0.4642 & 10 & Background radiation & 29 & \bf{UVBH0beta} \\
$\Delta z_{\rm H}$ & 0 & -5 & 0 & Background radiation & 30 & \bf{UVBH0Deltaz} \\
$\beta_{\rm He}$ & 1 & 0.25 & 4 & Background radiation & 31 & \bf{UVBHepbeta} \\
$\Delta z_{\rm He}$ & 0 & -1 & 0 & Background radiation & 32  & \bf{UVBHepDeltaz} \\
$t_0^*$ & 2.27 & 1.135 & 4.54 & Star-formation & 10 & MaxSfrTimescale \\
$q_{\rm eEOS}$ & 0.3 & 0.1 & 0.9 & ISM & 11 & FactorForSofterEQS \\
$b_{\rm IMF}$ & -2.3 & -2.8 & -1.8 & Stellar Evolution & 12 & IMFslope \\
$M_{\rm SNII,min}$ & 8 & 4 & 12 & Stellar Evolution & 13 & SNII\_MinMass\_Msun \\
$N_0$ [$10^{-3}$] & 1.3 & 0.85 & 1.75 & Stellar Evolution & 33 & \bf{SNIa\_Rate\_Norm} \\
$s$ & 1.12 & 0.88 & 1.36 & Stellar Evolution & 34 & \bf{SNIa\_Rate\_DTD\_power} \\
$\bar{e}_w=3.6A_{\rm SN1}$ & 3.6 & 0.9 & 14.4 & Galactic winds & 3 & WindEnergyIn1e51erg \\
$\kappa_{\rm w}=7.4A_{\rm SN2}$ & 7.4 & 3.7 & 14.8 & Galactic winds & 4 & VariableWindVelFactor \\
$\tau_w$ & 0.1 & 0.025 & 0.4 & Galactic winds & 14 & ThermalWindFraction \\
$\bar{m}_w$ & 0 & 0 & 4000 & Galactic winds & 15 & VariableWindSpecMomentum \\
$n_{\rm w,rec}$ & 0.05 & 0.005 & 0.5 & Galactic winds & 16 & WindFreeTravelDensFac \\
$v_{\rm w,min}$ & 350 & 150 & 550 & Galactic winds & 17 & MinWindVel \\
$f_{\rm w,Z}$ & 0.25 & 0.0625 & 1 & Galactic winds & 18 & WindEnergyReductionFactor \\
$Z_{\rm w,ref}$ [$10^{-3}$] & 2 & 0.5 & 8 & Galactic winds & 19 & WindEnergyReductionMetallicity \\
$\gamma_{\rm w,Z}$ & 2 & 1 & 3 & Galactic winds & 20 & WindEnergyReductionExponent \\
$1-\gamma_{\rm w}$ & 0.6 & 0.2 & 1 & Galactic winds & 21 & WindDumpFactor \\
$M_{\rm seed}$ [$10^{-5}$] & 8 & 2.5316 & 25.28 & BH growth & 22 & SeedBlackHoleMass \\
$A_{\rm Bondi}$ & 1 & 0.25 & 4 & BH growth & 23  & BlackHoleAccretionFactor \\
$A_{\rm Edd}$ & 1 & 0.1 & 10 & BH growth & 24 & BlackHoleEddingtonFactor \\
$A_{\rm AGN1}$ & 1 & 0.25 & 4 & AGN feedback & 5 & RadioFeedbackFactor \\
$f_{\rm re}=20A_{\rm AGN2}$ & 20 & 10 & 40 & AGN feedback & 6 & RadioFeedbackReiorientationFactor \\
$\epsilon_{\rm f,high}$ & 0.1 & 0.025 & 0.4 & AGN feedback & 25 & BlackHoleFeedbackFactor \\
$\epsilon_r$ & 0.2 & 0.05 & 0.8 & AGN feedback & 26 & BlackHoleRadiativeEfficiency \\
$\xi_0$ [$10^{-3}$] & 2 & 0.063 & 63.2 & AGN feedback & 27 & QuasarThreshold \\
$\beta_\chi$ & 2 & 0 & 4 & AGN feedback & 28 & QuasarThresholdPower \\
\bottomrule
\end{tabular}
\caption{A summary of the 35 parameters varied in the new $50\hMpc$ simulations. Note that some parameters are not varied symmetrically around the fiducial value, and some are varied in linear and some in logarithmic space. The parameter names and serial numbers from the public release data files\footnote{\url{https://camels.readthedocs.io/en/latest/parameters.html}} are provided for convenience in the last two columns. These also correspond to the names and serial numbers presented in \citet{NiY_23a}, except those in bold, which are those that are varied in this work for the first time. The first six parameters (serial numbers 1 through 6) are those that were varied in the original CAMELS simulations, i.e.~the LH set.}
\label{tab:parameters}
\end{table*}

\subsection{Simulation setup}
\label{sec:simulation_setup}

The initial conditions for all of the new $50\hMpc$ simulations are generated at a redshift of $z=127$ using second-order Lagrangian Perturbation Theory\footnote{\url{https://cosmo.nyu.edu/roman/2LPT/}} \citep{ScoccimarroR_98a,CrocceM_06a}. For every hydrodynamic simulation, which is initialized with $512^3$ collisionless dark matter particles and $512^3$ gas cells, we also run a corresponding N-body simulation that uses the same amplitudes and phases for the initial conditions and the same cosmological parameters (with a fixed gravitational softening as in the fiducial model). The hydrodynamic simulations further include $512^3$ gas tracers that allow Lagrangian analysis of the gas flow  \citep{GenelS_14a}. The mass resolution is $6.5\times10^7\hMsun\times(\Omega_m-\Omega_b)/0.251$ for dark matter particles and $1.27\times10^7\hMsun\times\Omega_b/0.049$ for baryonic particles in the hydrodynamical simulations, and $7.77\times10^7\hMsun\times\Omega_m/0.3$ for dark matter particles in the N-body simulations.

Each simulation produces 91 snapshots between $z=15$ and $z=0$. Halos and subhalos are identified on-the-fly using the Friends-of-Friends (FoF; \citealp{DavisM_85a}) and Subfind \citep{SpringelV_01} algorithms, and in post-processing by Rockstar \citep{BehrooziP_13f}. Merger trees are constructed using SubLink \citep{Rodriguez-GomezV_14a} on top of the Subfind catalogs and using ConsistentTrees \citep{BehrooziP_13e} on top of the Rockstar catalogs. The data formats of the snapshots, group catalogs, and merger trees all follow the previous generation of CAMELS-IllustrisTNG simulations, which is also essentially identical to that of the public data from the IllustrisTNG project\footnote{\url{https://www.tng-project.org}} \citep{NelsonD_19b}.

The new $50\hMpc$ simulations presented here are classified into three distinct sets, following previous CAMELS simulations, distinguished by the methodology used to sample the parameter space and by the choice of the initial random seed for the generation of the initial conditions, as follows:

\begin{itemize}
\item \textbf{SB35}. This set comprises 1,024 simulations, each characterized by a unique combination of the 35 cosmological and astrophysical parameter values, arranged in a Sobol sequence. The Sobol sequence is defined in a unit hypercube, which represents the parameter space via min-max normalization that is performed on each parameter, either in linear or logarithmic space, depending primarily on its dynamic range. These minimum and maximum values of the parameters were chosen to cover a wide range of physically plausible values, and they are identical to those used in \citet{NiY_23a} for the 28 common parameters. Furthermore, each simulation in this set employs a distinct random seed for the initial conditions so that it contains different large-scale structures.
\item \textbf{1P}. This set consists of four simulations per parameter, along with one fiducial simulation, where each simulation's parameter values differ from the fiducial case by just a single parameter, such that in this set the parameters are varied one at a time. For each parameter, the full range is divided (either in linear or logarithmic space, in correspondence with the respective choice for the SB35 set) into 4 equal segments, and the parameter values used for this set are the edges of these segments, one of which is always the fiducial, and two of which are, by construction, the minimum and maximum values. For most parameters, the fiducial value is located exactly in the middle of the range, such that the 1P simulations are symmetrical around it. However, in some cases, the fiducial value equals the minimum or maximum, resulting in variations that occur only on one side of the fiducial value. In some cases, the fiducial value is neither in the middle nor at the edge of the range but equals one of the midpoints. All of the 1P simulations utilize the same random seed for the initial conditions so that they contain the same large-scale structures, which facilitates direct, clean comparisons of the effects of single parameter variations on cosmic structures.
\item \textbf{CV}. This set is composed of 27 simulations that use identical cosmological and astrophysical parameter values, which are set to their fiducial values. They differ only in their initial conditions random seed. This set can facilitate studies of the impact of cosmic variance on the scatter between simulation results.
\end{itemize}

Finally, much of the results in this work are presented as a comparison and contrast between our new simulations, specifically the SB35 set, and our previous generation of CAMELS simulations. In particular, we focus the comparisons on the SB28 set of $(25\hMpc)^3$ boxes \citep{NiY_23a}, which were evolved with $2\times256^3$ resolution elements (implying the same mass resolution as our new simulations) and included variations in 28 parameters, which are a subset of the 35 parameters that are varied in the new SB35 set. We also make use of some comparisons to the LH set of $(25\hMpc)^3$ boxes, which has a similar setup but varies only six parameters (those with serial numbers 1-6 in \autoref{tab:parameters}), as well as to the CV set of the $(25\hMpc)^3$ boxes. All the simulation sets used in this paper are presented in \autoref{tab:simulations}.

\begin{table*}[ht]
\centering
\begin{tabular}{l|ccccc}
\hline
Simulation set & Volume & Number of particles & Number of simulations & Variations \\
 & $[(\hMpc)^3]$ & per simulation &  &  \\

\hline
SB35 & $50^3$ & $512^3$ & 1024 & Varying 35 parameters in a Sobol sequence \\
CV50 & $50^3$ & $512^3$ & 27 & Fiducial model, different initial conditions \\
1P & $50^3$ & $512^3$ & 141 & Varying 35 parameters one at a time \\
SB28 & $25^3$ & $256^3$ & 2048 & Varying 28 parameters in a Sobol sequence \\
CV25 & $25^3$ & $256^3$ & 27 & Fiducial model, different initial conditions \\
LH & $25^3$ & $256^3$ & 1000 & Varying 6 parameters in a Latin hypercube \\
ZoomGZ & halo zoom-ins & variable & 768 & Varying 28 parameters in a Sobol sequence \\
\hline
\end{tabular}
\caption{A brief description of the CAMELS simulation sets used in this work. The first three are the new second-generation CAMELS simulations that are introduced in this work, while the others are used for various comparisons.}
\label{tab:simulations}
\end{table*}

\section{The response to parameter variations}
\label{sec:variations}

In this section, we present an overview of the range of simulation outputs for various summary statistics, as seen in the different simulation sets from our new $50\hMpc$ simulations, along with a comparison to our previous set of simulations of $25\hMpc$ boxes.

\subsection{Global properties}
\label{sec:global_properties}

\begin{figure*}[ht!]
\includegraphics[width=0.99\textwidth]{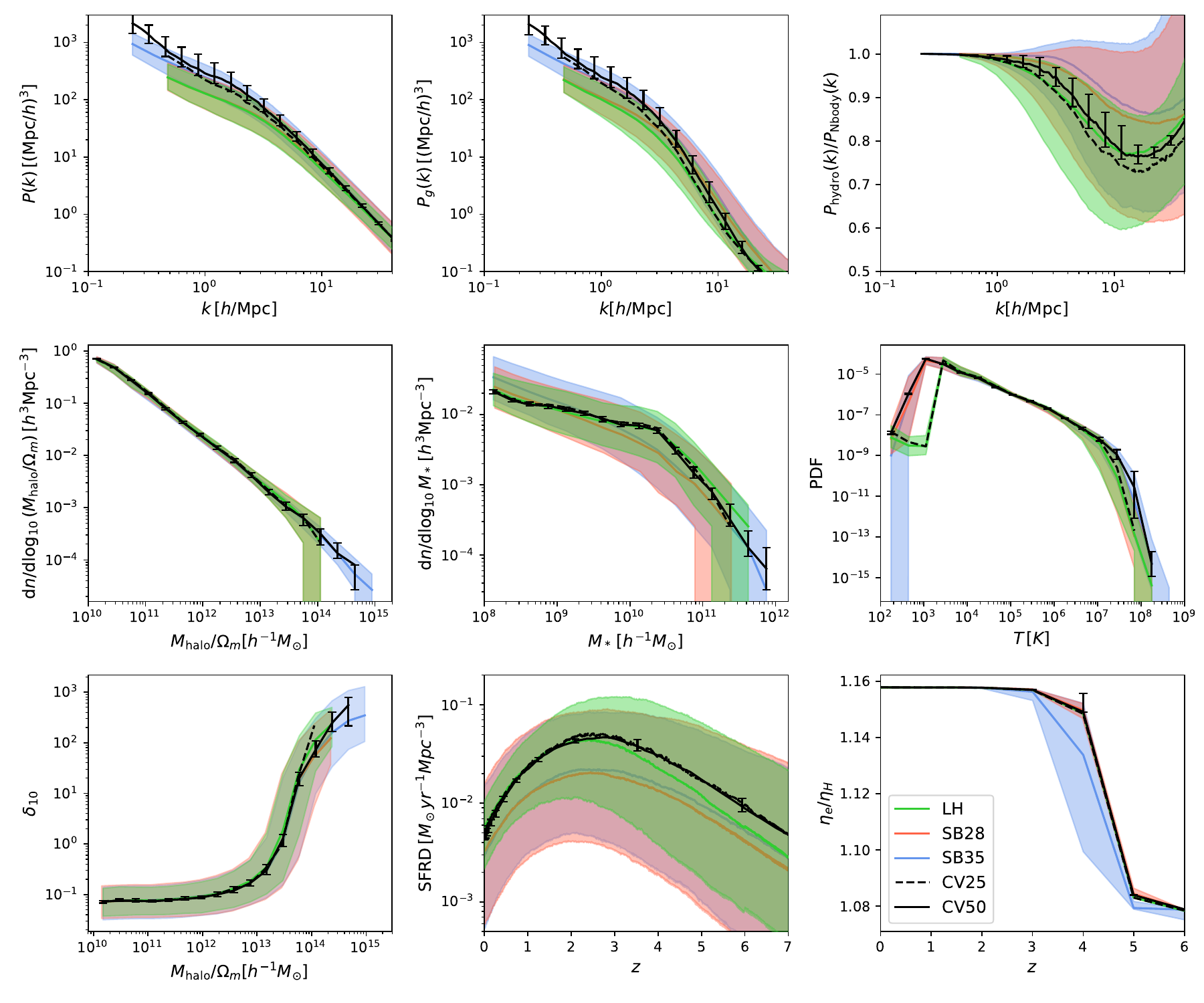}
\caption{Global statistical properties of the simulations across the SB35 (blue), SB28 (red), LH (green) and CV (black; solid for $50\hMpc$, dashed for $25\hMpc$) simulation sets, shown as medians (lines) and 16th-to-84th percentile ranges (shaded regions). The top row shows matter (left) and gas (middle) power spectra as well as the matter power spectrum ratio between corresponding hydrodynamical and N-body runs (right). The middle row displays halo mass functions (left), stellar mass functions (middle)\footnote{Note that the halo mass (left) is scaled by $\Omega_m$ since that considerably reduces the scatter between simulations, but no such scaling is applied for the stellar mass (middle), as the stellar mass function depends on a large number of parameters beyond $\Omega_m$.}, and subhalo inner (within the half-mass radius) temperature functions (right). The bottom row presents the galaxy overdensity-mass relation (left; \citealp{vandeVoortF_17a,SimsX_26a}), the cosmic star-formation rate density (middle), and the cosmic ionization history (right). SB35 extends coverage to larger scales and more massive halos, while SB28 and SB35 closely agree across overlapping ranges. Deviations in the LH set highlight the impact of differing parameter coverage. The scatter due to cosmic variance (shown only for CV50 for visual clarity) tends to be substantially smaller than in the other sets, which employ parameter variations.}
\label{fig:global_properties}
\end{figure*}

\autoref{fig:global_properties} presents nine facets of the results of our simulations, all quantities that summarize global properties of the simulation box: power spectra (and ratios thereof) in the top row, distribution functions of various halo properties in the middle row, and halo environment distribution, along with the redshift evolution of key properties of the simulation volume in the bottom row. In each panel, we compare our new SB35 set of simulations (blue) to two sets of our previous $25\hMpc$ simulations: the original LH set of CAMELS simulations that varies only 6 parameters (green) and the newer SB28 set (red) that varies 28 parameters. The solid curves show the median of all simulations in each set, while the shaded regions indicate the 16th-to-84th percentiles of the distributions among the individual simulation results (specifically, medians where relevant) in the respective sets. In addition, the CV sets of both box sizes are shown (black), with error bars that indicate the 16th-to-84th percentile scatter of the $50\hMpc$ CV set.

We first observe in \autoref{fig:global_properties} that, as expected, the SB35 set extends the results to larger scales (lower $k$ modes in the top row) and more massive objects (higher masses and temperatures in the middle row) compared to the other two sets, which have smaller volumes. This highlights the advantage of the new SB35 over the previous CAMELS simulations.

Further, we observe that SB35 and SB28 show very similar results (in their range of overlap) in many panels, while the 6-parameter LH set is discrepant. We interpret this as relating to SB35 and SB28 sampling a similar parameter space that is distinct from the one sampled by the LH set. Specifically, this highlights the limitations of the original CAMELS LH set, which covers only a small volume in the global parameter space of the IllustrisTNG model. The reason for these discrepancies is that the parameter variations included in any set tend to shift the results asymmetrically around the fiducial model; therefore, different parameter spaces result in differently shifted medians. A prominent example of this phenomenon is seen in the cosmic ionization history (bottom right), where all the sets show identical results except for SB35, in which variations of He reionization were added that cause a delay in the evolution of the ionization fraction \citep{VillasenorB_22a}. Another implication of this is seen in the low-$k$ regime of the power spectra (left and middle top panels), where the CV sets are at the edge, or even outside, the 16th-to-84th percentile ranges of the other sets. This stems from $\Omega_m$ variations (which are included in all sets except for the CV sets) leading to asymmetric modulations of the power spectrum on large scales, as shown explicitly using the 1P set in \citet{NiY_23a}.

Similarly, it is also interesting to note that the SB sets are shifted upwards compared to the CV and LH sets in terms of the power spectrum ratio between the hydrodynamical and N-body runs (top right), indicating lower feedback in the SB sets medians compared to the fiducial model, while the cosmic star-formation history (bottom middle) is shifted downwards, indicating stronger feedback compared to the fiducial model. These seemingly contradictory results stem from different forms of feedback, which operate on different (halo versus galaxy) scales, being responsible for these two distinct facets of the results.

The two CV sets, in each of which simulations only differ by their initial conditions but all use the fiducial parameter values, are largely indistinguishable between the two box sizes in their range of overlap, with the exception of the power spectra (top row), including the ratio of the matter power spectrum in the hydro simulations to that in the corresponding N-body simulations (top right). The larger boxes show more power at all scales $k\lesssim10h/\Mpc$, in particular for gas, and even the ratio between the power in the hydrodynamical and the N-body runs is elevated (at all scales $k\gtrsim1h/\Mpc$). This is a reflection of the enhanced mode coupling in the larger boxes. While the effect is non-negligible, it is smaller than the scatter among the runs with different parameter combinations of the SB and LH sets. Also, most statistics show a much smaller scatter between the simulations in the CV set than between those in the sets that vary parameters, as expected. A more detailed account of cosmic variance, as reflected through a comparison between the CV25 and CV50 sets, is provided in \autoref{sec:Pk}.

In Appendix \ref{sec:dependencies}, we present the detailed dependencies of two of the quantities discussed here, the power spectrum ratio (\autoref{fig:1P_Pk}) and the cosmic star-formation history (\autoref{fig:1P_SFRD}), on the 35 parameters that are varied in this work, utilizing the 1P set. The results demonstrate that individual parameters have, for the most part, distinct effects on these quantities; namely, if degeneracies exist, they are more complex and higher-dimensional than pairwise combinations between individual parameters.

\subsection{Galaxy scaling relations}
\label{sec:scaling_relations}

\begin{figure*}[ht!]
\includegraphics[width=0.99\textwidth]{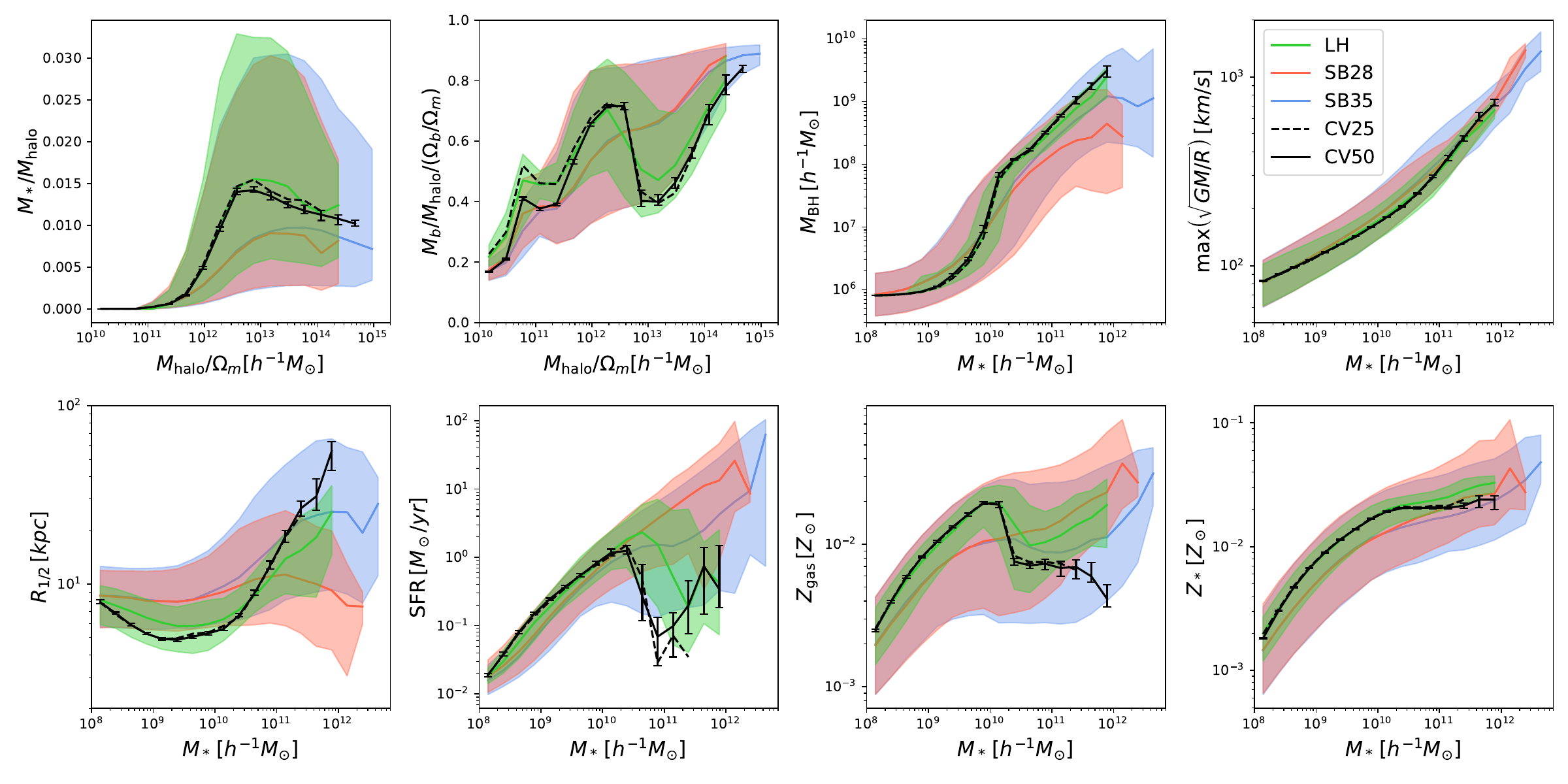}
\caption{Scaling relations between galaxy and halo properties across the SB35 (blue), SB28 (red), LH (green) and CV (black) simulation sets. The figure includes relations such as stellar mass, baryonic mass, black hole mass, circular velocity, stellar half-mass radius, star formation rate, and total gas metallicity (ISM and CGM combined) versus halo mass or stellar mass. The SB35 simulations extend the dynamic range toward higher masses, while the LH set shows an offset with respect to the SB sets (which largely agree with each other in their common well-sampled range of $M_\star\lesssim10^{11}\hMsun$) and a narrower scatter due to fewer varied parameters. The scatter due to cosmic variance (shown only for CV50 for visual clarity) tends to be substantially smaller than that due to the parameter variations in the other sets.}
\label{fig:galaxy_scaling_relations}
\end{figure*}

\autoref{fig:galaxy_scaling_relations} is analogous to \autoref{fig:global_properties}, except that it shows scaling relations between various halo or galaxy properties and their mass. Here too, we find that SB35 extends the coverage of the simulations towards more massive objects. SB35 and SB28 show similar results in their range of overlap, in contrast to the outlier LH set, which shows weaker feedback in low-mass galaxies (in the median) than the SB sets, as reflected in larger stellar and baryonic masses (top left), smaller stellar sizes (bottom left), and higher metallicities (bottom right). At high masses, the picture is partially reversed, with the SB sets showing smaller black hole masses, larger halo baryon masses, and a more extended star-formation main sequence. The LH set also has an appreciably smaller scatter due to the smaller number of varied parameters. Several galaxy properties (bottom row) have a larger scatter among the SB35 simulations than in the SB28 set, particularly at high masses of $M_\star\gtrsim10^{11}\Msun$, reflecting either the expanded parameter space or the more physical non-linear evolution of massive halos in these larger volumes (see further discussion in \autoref{sec:theory}).

The two CV sets produce very similar results to each other, with the largest difference being in the halo baryonic masses (top, second from left), where the baryonic masses of low-mass halos are somewhat suppressed in the larger boxes relative to the smaller boxes. This can potentially be a result of more stripping in the denser environments that exist in the larger boxes, but we reserve confirmation of this hypothesis for future work. We note that, similarly to \autoref{fig:global_properties}, the CV sets are not necessarily very close to the median of the other sets, which is due to the asymmetric effects of parameter variations. This is visible most notably in the SFR and total gas metallicity panels (middle bottom), where the fiducial CV model shows a strong break at $M_*\approx10^{10}\Msun$, which reflects sharp quenching, while the parameter-variation sets show less pronounced breaks. Finally, the spread in the CV50 set (namely, among the medians of the individual $50\hMpc$ boxes run with the fiducial model) is generally significantly smaller than that among the simulations in the other sets, where parameters are varied. This reflects the high statistical power of the CV50 set, which, in turn, reflects the small sample variance that is achieved with the $50\hMpc$ volumes in general, each containing thousands of galaxies. Given this large ratio in the spreads, we can expect it to be possible to extract the true dependencies of such scaling relations on simulation parameters without significant adverse effects due to sample variance ``noise". See \autoref{sec:Pk} for an extended discussion of cosmic variance as reflected through a comparison between the CV25 and CV50 sets.

In Appendix \ref{sec:dependencies}, we present the detailed dependencies of the stellar-to-halo mass ratio (closely related to the stellar mass function presented here; \autoref{fig:1P_MsMh}) on each of the individual 35 parameters that are varied in this work, utilizing the 1P set.

\subsection{The IGM response to varying the background radiation}
\label{sec:radiation}

At $z>2$, the balance between photoheating and adiabatic cooling sets the temperature of the cold diffuse IGM (gas with $T \lesssim 10^5 K$ and overdensities of $\Delta \lesssim 100$).
This leads to a power law relation in the temperature-density phase diagram of that gas, where $T = T_0\Delta^{\gamma -1}$, with $T_0$ as the characteristic temperature at the cosmic mean density and $\gamma$ as the slope that describes how the temperature scales with overdensity.
Scatter in this relation is due to heating from other mechanisms, such as galactic feedback.
Although this relation does not describe all the gas in the IGM, the evolution of $T_0$ and $\gamma$ can be used as summary statistics to study the thermal history and evolution of the IGM.
Here we examine the effects of varying four parameters that control the amplitude and timing of the background radiation model -- $\beta_\mathrm{H}$, $\Delta z_\mathrm{H}$, $\beta_\mathrm{He}$, and $\Delta z_\mathrm{He}$, which have not been varied in previous CAMELS simulations -- on these basic properties of the IGM, with a more comprehensive analysis reserved for future work. To do this, we fit the distribution of gas on the temperature-density plane in the regime mentioned above ($T \lesssim 10^5 K$ and $\Delta \lesssim 100$) with a power-law, from which we extract the values of $T_0$ and $\gamma$.

\begin{figure*}[ht!]
\includegraphics[width=0.99\textwidth]{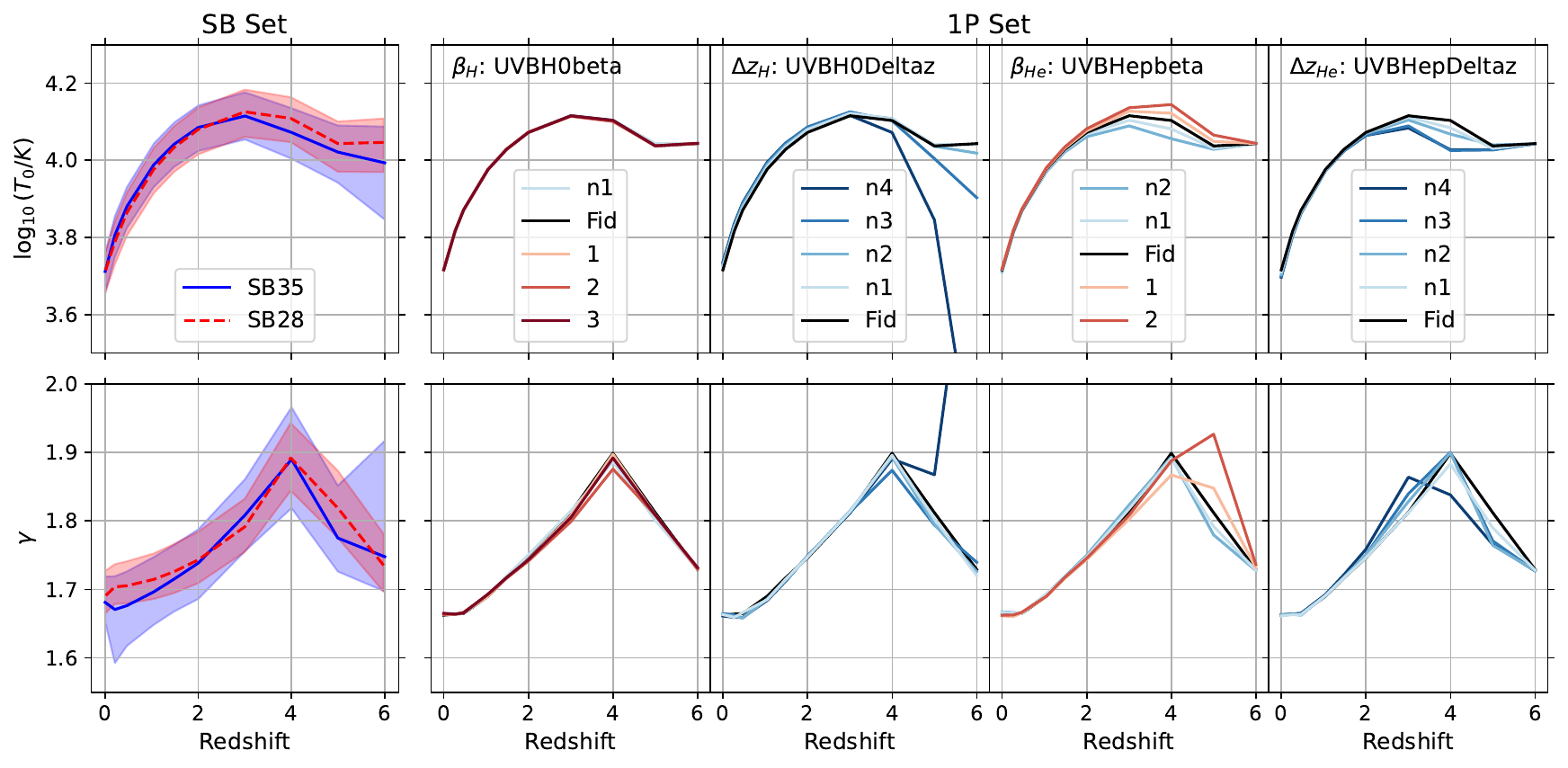}
\caption{The predicted $T_0$ (top) and $\gamma$ (bottom) from the $50\hMpc$ SB35 set (blue; left column), which is compared to the $25\hMpc$ SB28 set (red dashed; left column), as well as the $50\hMpc$ 1P set (right four columns). The $50\hMpc$ simulations vary the extragalactic ionizing background, which is not done in SB28. In the 1P panels, the black lines correspond to the fiducial simulation predictions, the blue and red lines correspond to decreases and increases in the varied parameter respectively. In the left column, shaded regions correspond to the 16th-to-84th percentile range of values. Greater variation is seen in these parameters closer to the end of the epoch of reionization, which also gives rise to more variation in the SB35 set compared to SB28.}
\label{fig:T0_gamma}
\end{figure*}

In \autoref{fig:T0_gamma}, we show the evolution of $T_0$ (top row) and $\gamma$ (bottom row).
The right four columns show the resulting $T_0$ and $\gamma$ for the 1P set variations of the background radiation parameters.
Changes due to variations in $\beta_H$ and $\Delta z_H$ should be expected closer to H reionization ($z>5$), while changes due to $\beta_{He}$ and $\Delta z_{He}$ are expected closer to He reionization ($z<5$). The variation ranges for these four parameters (see Table \ref{tab:parameters}) were chosen such that they largely encompass the difference in the amplitude of the radiation fields at any given cosmic time between the \citet{FaucherGiguereC_09a} background, which is used in the fiducial IllustrisTNG model, and the more recently derived \citet{PuchweinE_19a} background.

$\beta_H$ scales the photoionization and photoheating rates of H for the background radiation model.
Minimal to no change is seen in $T_0$ and $\gamma$ for variations in this parameter; however, this is not indicative of what is found for other statistics probing the IGM.
For example, variations in the Lyman-$\alpha$ forest 1D transmitted flux power spectrum, which directly probes diffuse neutral hydrogen in the IGM, can be observed for variations in this parameter, as demonstrated in \citet{VillasenorB_22a}.
Changes due to $\Delta z_H$ occur at $z>4$ as this value shifts the end of H reionization to later times, resulting in delayed heating and, therefore, lower values of $\gamma$, as expected \citep{WellsA_24a}.
This parameter also gives rise to an upward turn in $\gamma$ at $z=5$ when its value is at its extreme (the n4 case), as can be seen in the middle bottom panel.
However, treating the temperature--density distribution as a power law relation does not make sense prior to the epoch of reionization, so we regard this regime as spurious.

$\beta_{He}$ scales the photoionizing and photoheating rates of He, and $\Delta z_{He}$ shifts the end of He reionization to later times.
The impact of these parameters is more clearly seen in the redshift range that \autoref{fig:T0_gamma} covers, as it is closer to the epoch of He reionization \citep{SokasianA_02a,GnedinN_22a,VillasenorB_22a}. 
Scaling $\beta_{He}$ up increases $T_0$ and vice versa.
Changes in $\gamma$ when varying $\beta_{He}$ are not intuitive, as a change in the normalization of the ionizing background alone should not necessarily affect $\gamma$. 
These changes are confined to $z=4-5$, a period of time during which the photoionization and photoheating rates of HeII increase substantially in the employed \citet{FaucherGiguereC_09a} background radiation model.
In this way, increasing or decreasing $\beta_{He}$ changes the redshift at which the HeII ionizing background first becomes strong enough to have a measurable effect.
Similarly, shifting the end of He reionization to later times (decreasing $\Delta z_{He}$) decreases $T_0$ and shifts $\gamma$.
In general, the range of measured values for $T_0$ and $\gamma$ aligns with what we might expect from the implemented parameter variations.

For the SB sets, shown in the left column in \autoref{fig:T0_gamma}, we compare the median results from SB35 (blue solid lines), where the background radiation parameters are varied, to those from SB28 (dashed red lines), where they are not.
The shaded regions represent the 16th-to-84th percentile range of values from each simulation set.
The median results of the two simulation sets differ on a scale that mirrors the responses to the individual parameters based on the 1P set discussed above, and the variations within the SB35 set are greater than those of the SB28 set due to the inclusion of the additional parameters that vary the background radiation, as expected.
Meanwhile, the slight upturn in $\gamma$ seen for SB35 at $z\gtrsim5$ originates from variations of $\Delta z_{\rm H}$, which we regard as spurious in this redshift range, as discussed above.

\section{Parameter inference}
\label{sec:inference}

\subsection{Theoretical expectations}
\label{sec:theory}

In this section, we showcase a few applications of parameter inference from SB35 simulation outputs that, with a careful comparison to results from the SB28 set, demonstrate where the new $50\hMpc$ simulations provide advantages over the previous $25\hMpc$ simulations. We begin, in this subsection, with a theoretical discussion of differences we might expect between the constraining power of these two simulation suites when performing parameter inference, a discussion that will guide the interpretation of the results covered in the following subsections.

The most na\"ive expectation would be for an improvement in constraining power, namely a narrowing of posterior variances, that scales with the increase in the simulation volume used for the inference; namely, in our case a factor of $\sqrt{8}$ decrease in posterior widths from SB28 to SB35. The reason for this is that the initial conditions of the simulations contain almost exactly eight times more information -- specifically, more modes of the initial Gaussian random field -- in our larger volumes compared with the smaller volumes. However, several factors add nuance to this zeroth-order expectation.

On one hand, we might expect the improvement to scale more strongly because not only has the number of modes that are resolved in the $25\hMpc$ boxes, those with $k[h/\Mpc]>0.36$, increased eightfold in the $50\hMpc$ boxes, but additional modes with a smaller wavenumber $0.18<k[h/\Mpc]<0.36$ are present in the $50\hMpc$ boxes that cannot be resolved in the smaller volumes. These modes, even though there are very few of them, contain additional information, to different degrees for different parameters \citep{TegmarkM_97a}.
On the other hand, two types of effects may cause degradation in the na\"ive expectation for the scaling.

First, effects related to non-linear structure formation (sourced both by gravity and baryonic physics) are expected to reduce the amount of available information via mode coupling \citep{TegmarkM_97a}.
The basic phenomenon is that since all the modes in our relatively small boxes become correlated with one another by $z=0$, the amount of information they contain does not scale linearly with the number of modes but is instead weaker than that. In other words, the effective number of independent modes decreases with cosmic time and does not scale linearly with volume in the late-time, non-linear evolution regime.
Further, the mode coupling itself is stronger in our larger boxes due to the existence of additional low-k modes that are not present in the smaller boxes, as well as due to the smaller suppression of the non-linear growth of the modes, which occurs because the size of the box itself is not far from the wavelength of the non-linear modes. This suppression, which implies a non-physical suppression of structure formation, is more pronounced in the $25\hMpc$ boxes, leading to suppressed mode coupling there. There is a subtle interplay here between the correctness of the physics and the information content; due to both of these considerations (additional low-k modes and the fact that the fundamental mode in the larger boxes is farther from the non-linear scale than that of the smaller boxes), structure in the larger boxes evolves with higher physical fidelity, but since both lead to stronger mode coupling, that higher physical fidelity is associated with a degree of information loss.

\begin{figure}[ht!]
\includegraphics[width=0.46\textwidth]{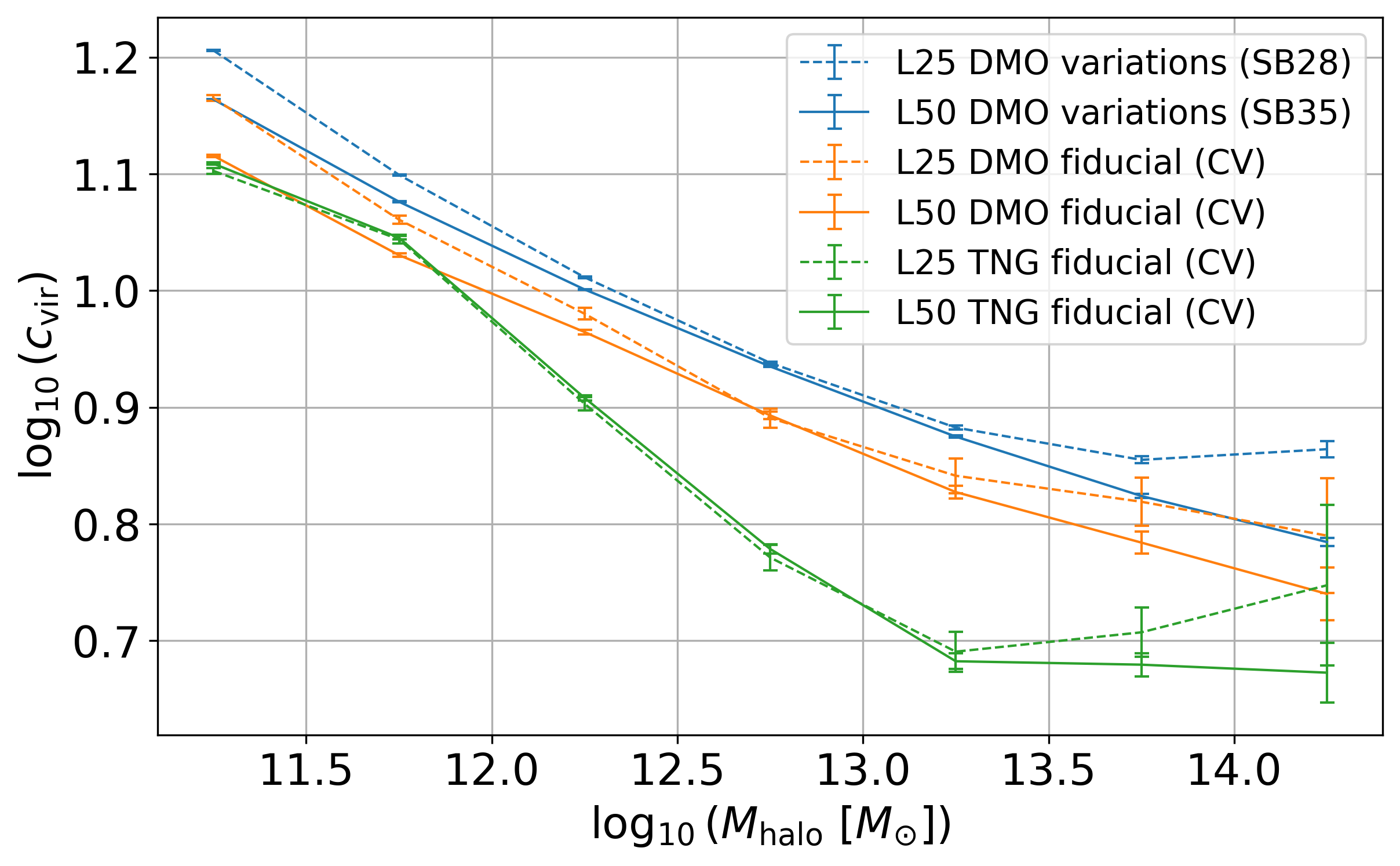}
\caption{A comparison of the concentration--mass relation for dark matter halos between our $50\hMpc$ and $25\hMpc$ boxes. We compare three pairs of simulation sets: hydrodynamical with the fiducial TNG model (CV sets; green), pure N-body with the fiducial cosmological parameters (CV sets; orange), and pure N-body with variations in five cosmological parameters (SB28 and SB35 sets; blue). In all cases, the lack of larger-scale modes in the $25\hMpc$ boxes leads to an upwards bias in their resulting concentration-mass relation. The physical reason for this upwards bias is that late-time, non-linear structure formation is suppressed in these smaller volumes such that halos form artificially earlier, leading (by way of the usual anti-correlation between concentration and formation time) to higher concentrations. This indirectly suggests that mode coupling and its associated information loss are likely also suppressed in these small boxes. Our CV sets comprise 27 simulations each, leading to marginal statistical significance of the effect at high masses (which indeed could not be detected by \citet{PowerC_06a}), but our SB sets (blue) comprise more than a thousand simulations each, leading to a very significant signal.}
\label{fig:c_vs_Mvir}
\end{figure}

Evidence for these aspects of non-linear growth can be indirectly seen in \autoref{fig:c_vs_Mvir}, where we demonstrate, through the concentration--mass relation \citep{LudlowA_14a,IshiyamaT_21a,SoriniD_25a}, that the evolution of dark matter halos is systematically different in our two box sizes.
As discussed above, the non-linear matter power spectrum is also not yet converged on the scales relevant to our simulations (see \autoref{fig:global_properties}), and this has been shown to be the case even with boxes of a side length of $128\hMpc$ or larger \citep{SchneiderA_16a}.
An alternative way to frame this effect is that while the sample (cosmic) variance in the smaller boxes is larger than in the larger boxes due to their sampling of a smaller number of modes, it is, in fact, suppressed compared to $25\hMpc$ sub-volumes (octants) from the $50\hMpc$ boxes due to the suppressed mode coupling. This is another way to think of the weaker-than-na\"ive scaling of the improvement in parameter constraints as related to an improvement in the physical fidelity of the simulations as the volume is increased.

Second, effects related to the parameter space may present as a weaker-than-na\"ive scaling. One such effect is that our SB35 set includes variations of 7 additional parameters with respect to the SB28 set. If there exist (even partial) degeneracies in the response of the simulation results between these parameters and the 28 parameters that are in common, constraints on these 28 original parameters may not shrink as much as they otherwise would. In this sense too, then, a weaker-than-na\"ive scaling of the improvement on parameter constraints is related to an improvement in the simulation setup -- namely, sampling more parameters. However, even if that is not the case and the additional 7 parameters have no impact on the data used for the inference, improvements may not be reflected in the marginal posteriors of individual parameters. This is because of potential degeneracies in this high-dimensional parameter space. While in the Gaussian case, as can be seen straightforwardly in the Fisher formalism, the marginals of the posterior shrink along with the covariance (which ideally scales inversely with the volume, as discussed above), for a general posterior, with curvature or multi-modalities, that is not necessarily the case. The posterior volume may shrink due to the increased simulation volume, but its marginals may shrink by much less. This effect can, in principle, be detected by inferring the full posterior rather than just its marginals; however, that may be technically challenging in a 35-dimensional space, and in this work, we limit ourselves to inferring posterior marginals.

In \autoref{sec:Pk} through \autoref{sec:graphs} we focus on comparisons of inference on two cosmological parameters: $\Omega_m$ and $\sigma_8$. It is worth noting that the Fisher information in the linear approximation gains more from the additional modes that are present in the $50\hMpc$ boxes but not in the $25\hMpc$ boxes for $\sigma_8$ than for $\Omega_m$, because the power spectrum in those additional modes, $k\sim0.2-0.3h\Mpc^{-1}$, is more sensitive to the former than the latter. This may hint that we might expect better improvements for $\sigma_8$ than for $\Omega_m$; however, there are several other confounding factors, as discussed above.

\subsection{Matter power spectra}
\label{sec:Pk}

We open the series of parameter inference exercises with a simple summary statistic that characterizes the distribution of matter in the simulation volumes: the non-linear matter power spectrum. We split the simulations on a 75\%-15\%-15\% basis into training, validation, and test sets, and train fully connected neural networks using the \textsc{ltu-ili} package \citep{HoM_24a} to infer the posteriors in the $\Omega_m$-$\sigma_8$ space using normalizing flows, while optimizing their hyperparameters using \textsc{Optuna} \citep{Optuna}. The input data are the $z=0$ non-linear total matter power spectra, as calculated with \textsc{pylians}\footnote{\url{https://pylians3.readthedocs.io}} \citep{Pylians}. As a sanity check, we also train a random forest regressor with the same inputs and point estimates of the parameter values as outputs, and obtain very similar results to the means of the posteriors we obtain with the normalizing flows. For the SB28 $25\hMpc$ boxes, we use only 1,024 simulations in order to match the size of the SB35 simulation suite, ensuring that the size of the training set does not play a role in any differences in the results. The utilized data for SB28 include 198 k-bins in the range $0.36<k[h/\Mpc]<50$, while for the SB35 $50\hMpc$ boxes, we have 397 bins in the range $0.18<k[h/\Mpc]<50$. We have verified that using rebinning to compress the power spectra of SB35 to 198 bins has no discernible effect on the results, as expected. We limit the data to $k[h/\Mpc]<50$ to roughly match the Nyquist frequency of the grids we use to calculate the power spectra (with $512^3$ grid cells for the $25\hMpc$ boxes and $1024^3$ grid cells for the $50\hMpc$ boxes).

\begin{figure}[ht!]
\includegraphics[width=0.49\textwidth]{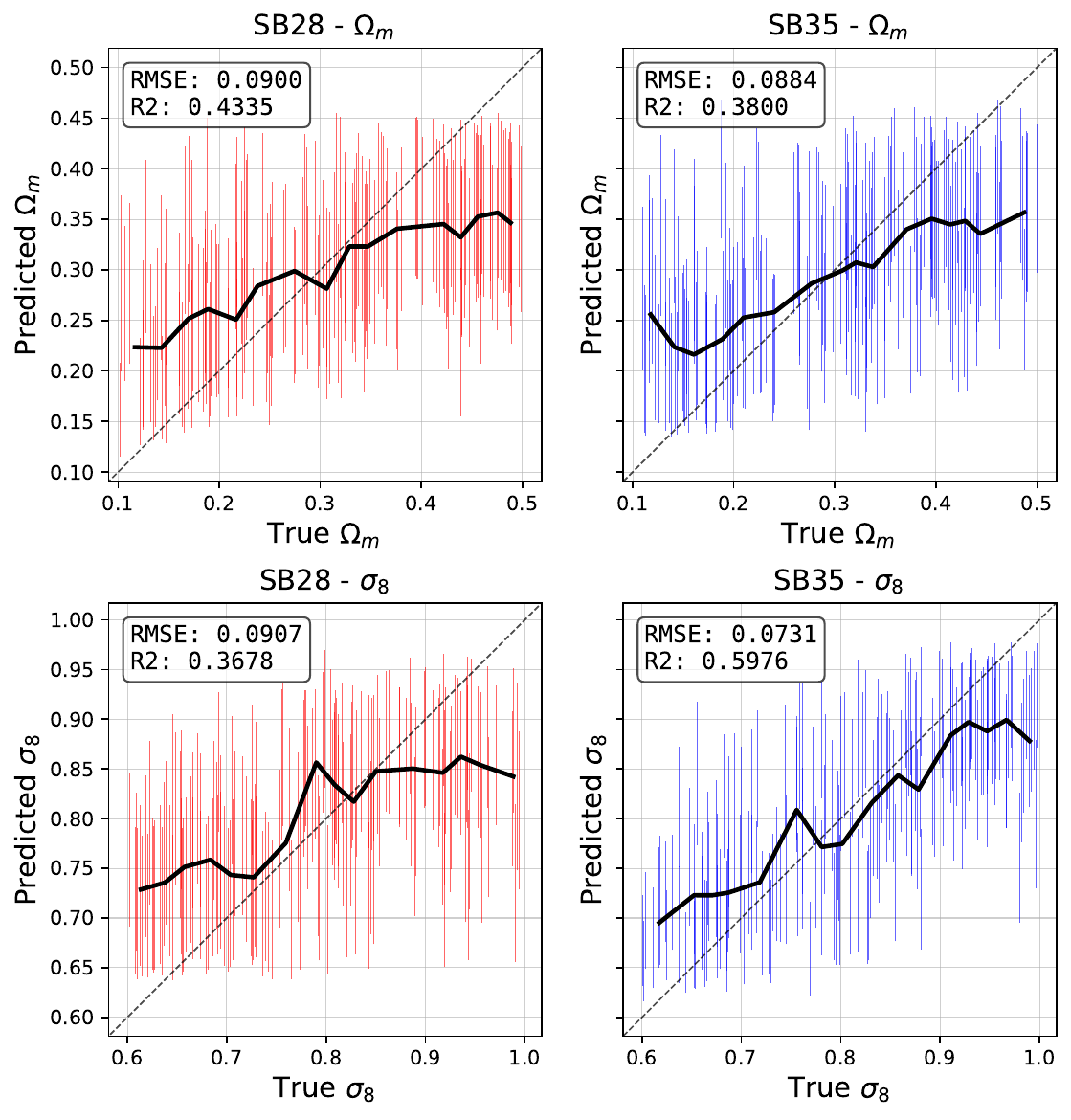}
\caption{Comparison of parameter inference from the $z=0$ non-linear matter power spectrum in SB28 (left panels) and SB35 (right panels). Each panel shows predicted versus true values for the tested parameters ($\Omega_m$ at the top, $\sigma_8$ at the bottom), with black solid curves indicating the running medians. The larger volume of SB35 leads to somewhat tighter predictions, especially for $\sigma_8$, but the RMSE ratios are far from fully reflecting the eightfold larger volumes of the SB35 simulations. We interpret the information loss as a consequence of mode coupling.}
\label{fig:inference_Pk}
\end{figure}

We present the results from this exercise in \autoref{fig:inference_Pk}, where the SB28 results are on the left and the SB35 results are on the right. We find that the root-mean-squared error (RMSE) value of $\approx0.09$ for $\Omega_m$ (top) barely improves between the smaller SB28 and the larger SB35 boxes. For $\sigma_8$ (bottom), the RMSE improvement is a factor of $\approx1.3$, which is much smaller than the na\"ive scaling with the volume ratio, $\sqrt{(50/25)^3}\approx2.8$. We repeated the inference using SB35 while dropping the two lowest-k bins, the ones that cannot be probed with SB28 due to the fundamental mode of these $25\hMpc$ boxes, finding essentially identical results. This indicates that in this case the improvement on $\sigma_8$ constraints is driven by the smaller noise (due to smaller sample variance) in the $k$-bins that are common between the two volumes, which are the vast majority, $\approx99.5\%$, of the bins. It is worth noting, however, that as the box size is increased further in future simulations, one can expect the additional large-scale modes to eventually contribute more significantly.

\begin{figure}[ht!]
\includegraphics[width=0.49\textwidth]{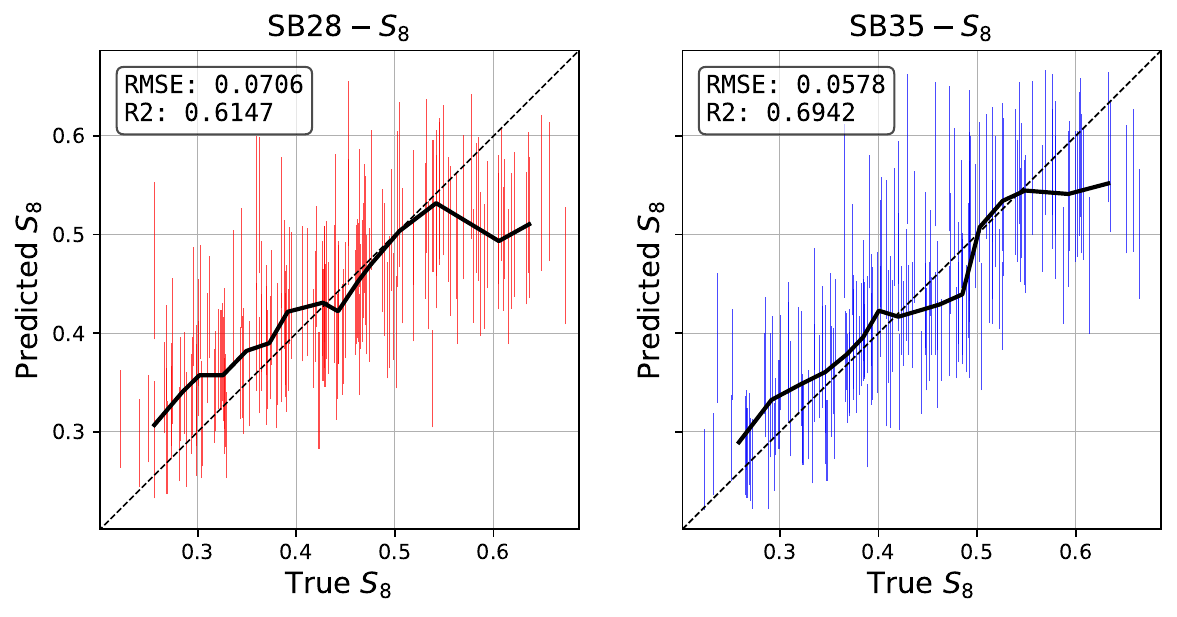}
\caption{Comparison between SB28 (left) and SB35 (right) for $S_8$ parameter inference from the matter power spectrum, in the same format as \autoref{fig:inference_Pk}. The tightening of the constraints with volume is not substantially stronger than that obtained for $\Omega_m$ and $\sigma_8$ individually, suggesting that degeneracies are likely not the dominant factor in the weak volume scaling.}
\label{fig:inference_Pk_S8}
\end{figure}

Why is it, then, that the constraints from the power spectrum barely become tighter as the cosmological volume (and accordingly, the data volume) is increased between SB28 and SB35? The results shown in the appendix (\autoref{fig:1P_Pk}) indicate that the additional 7 parameters that are varied in SB35 but not in SB28 have a minimal effect on the power spectrum. This suggests that these parameters likely do not play a role in reducing the constraining power of the power spectrum on $\Omega_m$ and $\sigma_8$\footnote{We note, however, that our constraints for $\Omega_m$, which correspond to a relative accuracy of $\sim30\%$, are less tight than the relative accuracy of $\sim20\%$ reported in \citet{Villaescusa-NavarroF_21c} using the power spectra of the LH set, namely the original CAMELS simulations that varied only six parameters. This difference most likely stems from more significant degeneracies in the 28-dimensional parameter space probed here compared to the original 6-dimensional space of the LH set.}. As discussed in \autoref{sec:theory}, it is, in principle, possible for the posterior to shrink along directions that correspond to certain parameter combinations more than its marginals. To probe this possibility, we train a network to infer $S_8\propto\sigma_8\sqrt{\Omega_m}$. The results, as shown in \autoref{fig:inference_Pk_S8}, do not suggest that such an effect is in play; while the constraints on $S_8$ are tighter than on $\Omega_m$ and $\sigma_8$ individually, they tighten from SB28 to SB35 merely by a factor of $\approx1.2$, comparably to the volume scaling of the individual parameters.

\begin{figure}[ht!]
\includegraphics[width=0.49\textwidth]{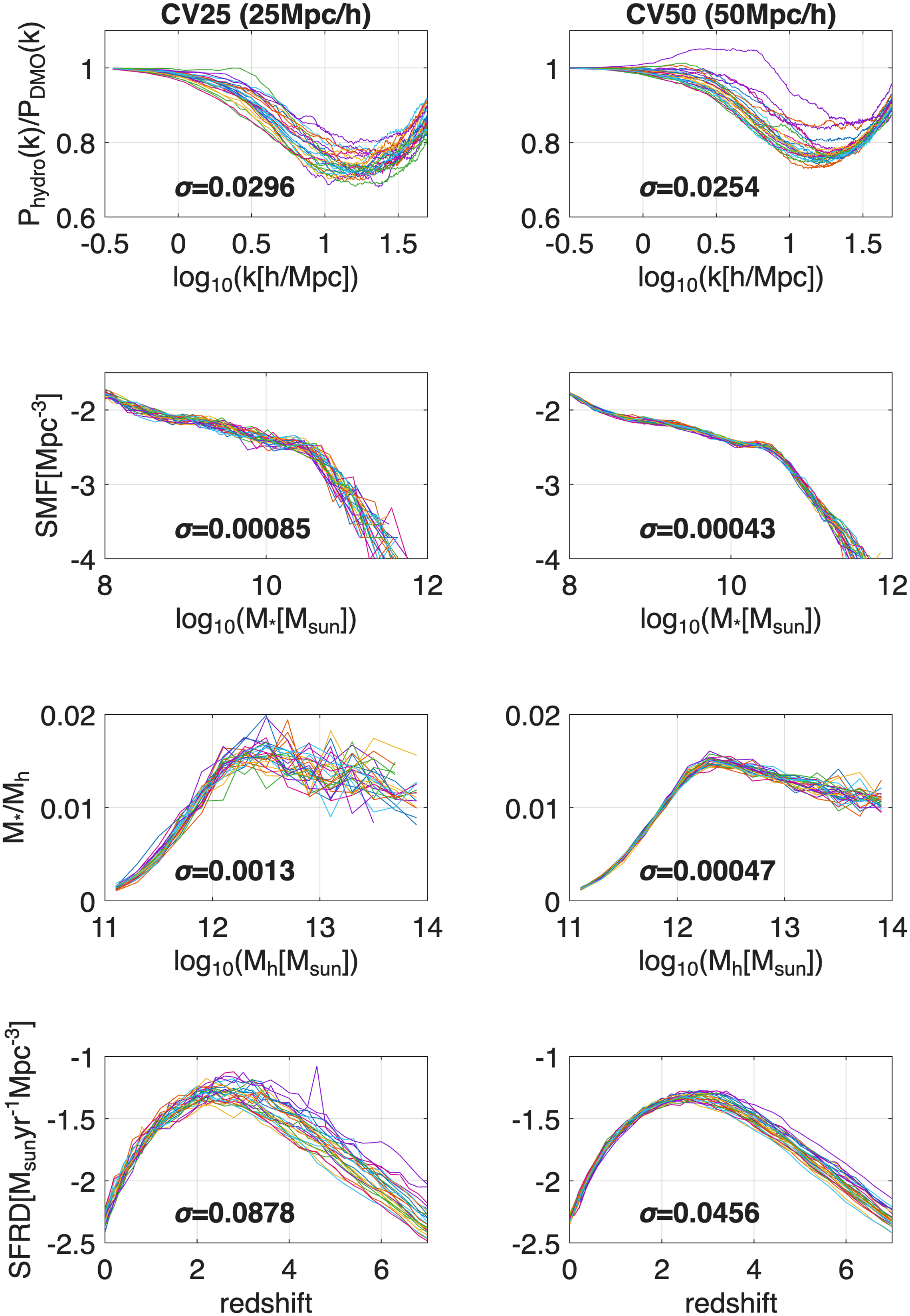}
\caption{A comparison of the cosmic variance between $(25\hMpc)^3$ and $(50\hMpc)^3$ volumes using our CV25 (left) and CV50 (right) sets run with the fiducial IllustrisTNG model. Four quantities are displayed, from top to bottom: the matter power spectrum ratio between the hydrodynamical and their corresponding dark matter-only simulations, the stellar mass function, the stellar-to-halo mass ratio, and the history of cosmic star-formation density. Each panel displays the scatter among the 27 curves and indicates the standard deviation between them (averaged over the x-axis bins). It is immediately clear that the cosmic variance scales with volume differently for different quantities. In particular, the power spectrum ratio decreases by a factor of only $\approx1.15$ between $(25\hMpc)^3$ and $(50\hMpc)^3$ volumes, while the other quantities scale more similarly to the `ideal' scaling factor of $\sqrt{(50/25)^3}\approx2.8$.}
\label{fig:CV}
\end{figure}

Instead, we argue that another factor plays a significant role in the weak improvement of the constraints with volume: cosmic variance. To infer the value of a parameter, the dependence of the data vector on the parameter must be distinguishable from, namely comparable to or larger than, the scatter of the data between different samples, which we refer to here as cosmic variance. In \autoref{fig:CV} (top row) we demonstrate that, indeed, the scatter between the power spectrum ratio in the different boxes in our CV sets, which all use the same fiducial IllustrisTNG model but differ in their initial conditions, is only $\approx15\%$ smaller in CV50 relative to CV25. If the modes in the different octants of the CV50 boxes were independent of each other, we would expect the scatter in each CV50 box to be smaller by a factor of $\sqrt{(50/25)^3}\approx2.8$ than in an octant, and similarly relative to a CV25 box. This would then translate approximately to the same factor of improvement for the parameter constraints, so long as that level of cosmic variance is smaller than the dependence of the power spectrum ratio on the parameters.
The fact that the cosmic variance is similar between the two volumes is interpreted as a consequence of strong mode coupling on the scales that are simulated here, and as the cause for the minuscule shrinking of the posteriors on the parameters.

For reference, \autoref{fig:CV} also displays the cosmic variance comparison between the two sets for three additional quantities: the stellar mass function, the stellar-to-halo mass ratio, and the history of cosmic star-formation density. Of these three, the two that are sensitive directly to the halo mass function, namely the stellar mass function (second row) and the cosmic star-formation density (bottom row), scale by a factor of $\approx2$ between the volumes. This much closer scaling to `ideal' than the power spectrum ratio reflects that the mass functions in different octants of the CV50 boxes are not quite as coupled to each other as the amplitudes of the power spectrum modes. Once the halo mass function is taken out by considering the stellar-to-halo mass ratio (third row), the scaling becomes essentially the `ideal' one: $0.0013/0.00047=2.77\approx\sqrt{(50/25)^3}$. This suggests that the internal properties of halos, at least concerning their stellar content, are essentially independent of each other in sub-volumes at these scales, such that increasing the volume leads to averaging and hence to lower scatter.

\subsection{Projected maps}
\label{sec:projected_maps}

Here we study the constraining power of another, much more expressive measure of the spatial distribution or clustering of matter on the same two cosmological parameters, $\Omega_m$ and $\sigma_8$: two-dimensional maps of $z=0$ projected total matter density in $5\hMpc$-thick slices\footnote{The code used to generate the results in this section is available at \url{https://github.com/YongseokJo/CAMELS_SB35_comparison_to_SB28}.}. For a meaningful and fair comparison between our various simulation boxes, we keep the thickness fixed and include three types of maps: i) $(25\hMpc)^2$ maps of the full field from SB28 simulations, ii) $(50\hMpc)^2$ maps of the full field from SB35 simulations, and iii) $(25\hMpc)^2$ maps that are single-quadrant cutouts of SB35 slices. For each of these three cases, we train and test a network to infer the values of these two parameters, and we also perform cross-tests between the two types of $(25\hMpc)^2$ maps. We perform these experiments for normalized maps as well as for unnormalized maps that retain the information on the monopole, which in this case is simply the total mass contained in the slice from which the map is made.

Each SB28 (SB35) simulation produces 15 (30) slices: 5 (10) adjacent slices projected along each of the cartesian axes.
For our main analysis, we use 15,360 maps for each set, which are taken from 1,024 simulations with exactly 5 non-overlapping slices projected along each of the 3 spatial dimensions. For this analysis, we use the exact same 70\%/15\%/15\% split of the simulations into training, validation, and test sets as in the previous subsection to allow a like-for-like comparison. However, we also test the effect of doubling the data volume and report on that at the end of this subsection; this is achieved for the SB28 set by using all 2,048 simulations that comprise it, and for the SB35 set by using all 10 available non-overlapping slices per dimension.

These two-dimensional projected maps are generated by extracting from the simulation data the spatial coordinates and physical properties of particles relevant to a chosen field; in our case here, the mass.
Each gas and dark matter particle is assigned an effective radius $R$ defined as the distance to the 32nd nearest particle of the same type. 
For star and black hole particles, we set $R=0$, since the resolution of our two-dimensional maps is too coarse to capture internal galactic structure; in these cases, the window function reduces to a Dirac delta.
Each particle is assigned a window function $W_{2D}(r,\theta)$ on the projection plane, obtained by integrating the three-dimensional top-hat kernel
\begin{equation}
W_\mathrm{3D}(r,\theta,z) = 
\left\{ 
\begin{array}{cc} 
\frac{3}{4\pi R^3} & \mbox{if }r<R \\
0 & \mbox{otherwise}
\end{array}\right.,
\end{equation}
namely $W_\mathrm{2D}(r,\theta) = \int_z W_\mathrm{3D}(r,\theta,z)\mathrm{d}z$.
Particle attributes are then deposited onto a $256 \times 256$ ($512 \times 512$) grid (for SB28 and SB35, respectively), i.e.~using a grid resolution of $\approx98\hkpc$, following \citet{Villaescusa-NavarroF_22c}. 
The procedure involves uniformly sampling 1000 tracers within each particle radius, subsequently assigning each tracer to the corresponding pixel. 
This numerical approximation method asymptotically approaches the precise outcome as the number of tracers increases.

We employ a convolutional neural network (CNN) to perform parameter inference from the projected maps.
The network architecture is built from a deep sequence of convolutional layers to respect the symmetries of the simulation boxes, using zero padding. 
Each convolutional block is followed by batch normalization and a LeakyReLU activation, while the resolution is progressively reduced through strided convolutions from $256 \times 256$ ($512 \times 512$) down to $1 \times 1$. 
At the same time, the number of channels grows layer by layer, controlled by a hyperparameter $H$, such that deeper layers capture increasingly complex features.
The resulting feature maps are flattened and passed through fully connected layers with dropout regularization before producing the outputs \citep[refer to Sec. 3.1.1 of][for the detailed architecture]{Villaescusa-NavarroF_22c}.
The network predicts both the mean and the standard deviation of the marginal posterior for the cosmological parameters $\Omega_m$ and $\sigma_8$, using the \citet{JeffreyN_20a} loss function.

We present the main results of this subsection in Table \ref{tab:maps_RMSE}, which provides the mean (over individual maps) RMSE of the various experiments normalized by the mean RMSE values of the networks that are trained and tested on SB28 no-monopole (normalized) maps. The latter are $0.04\pm0.0034$ for $\Omega_m$ and $0.065\pm0.008$ for $\sigma_8$\footnote{We note that, analogously to the case of the power spectrum, these constraints on $\sigma_8$, which correspond to a relative accuracy of $\sim9\%$, are less tight than the relative accuracy of $\sim2.4\%$ reported in \citet{Villaescusa-NavarroF_21c} using maps of the LH set, namely the original CAMELS simulations that varied only six parameters. This difference most likely stems from more significant degeneracies in the 28-dimensional parameter space probed here compared to the original 6-dimensional space of the LH set.}, where the quoted errors represent the aleatoric uncertainties, estimated by the standard deviation of the mean RMSE over five random $50\%$ sub-samples of the full data set (namely, 1,024 simulations out of the full 2,048 for SB28, and 5 slices per dimension per box out of the full 10 slices for SB35). We do not calculate epistemic uncertainties.

\begin{figure*}[ht!]
\includegraphics[width=0.99\textwidth]{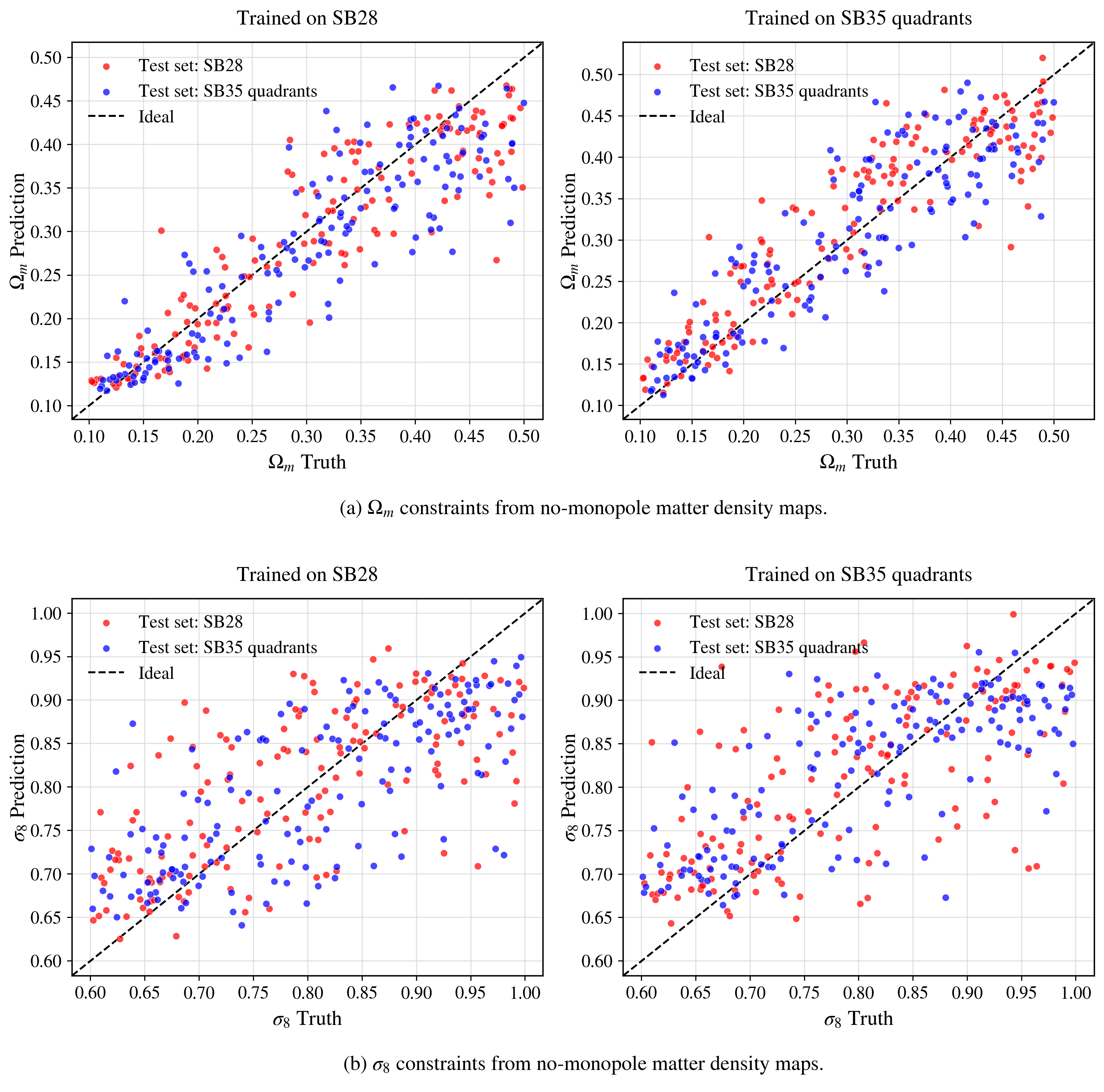}
\caption{Parameter inference for $\Omega_m$ (top) and $\sigma_8$ (bottom) using no-monopole $25\hMpc$ maps from SB28 and SB35 (quadrants), including both self- and cross-validations. Models trained on SB28 maps (left) and on SB35 quadrants (right) are tested on SB28 maps (red) and on SB35 quadrants (blue). No clear out-of-distribution biases are observed.}
\label{fig:maps}
\end{figure*}

We begin by discussing the results from no-monopole maps, for which the relative RMSEs are given in the left and middle sections of Table \ref{tab:maps_RMSE}, and a detailed view is presented in \autoref{fig:maps}. First, we observe that all tests with $(25\hMpc)^2$-sized maps, whether full-field from SB28 or quadrants from SB35, and whether tested in-distribution or across the two types, perform similarly well. As the RMSEs in Table \ref{tab:maps_RMSE} are normalized to the SB28-SB28 case, we see a value of $1.0$ for that case (by construction), with deviations from unity being on the order of $10\%$ for the other cases, which are comparable to the aleatoric uncertainty reported above. This can also be seen in \autoref{fig:maps}, where all four combinations of training and test sets (for each parameter) display similar distributions, and no clear signs of out-of-distribution biases are present. This implies that the amount of information on these parameters is similar, regardless of the presence of additional low-k modes in the SB35 quadrant maps and despite SB35 incorporating variations of 7 additional parameters relative to SB28 maps. There is a hint that the additional low-k modes play different roles for the two parameters: SB35 quadrants perform slightly worse for $\Omega_m$ but slightly better for $\sigma_8$ compared to SB28 maps (with normalized RMSEs of 1.03 and 0.96, respectively). This may be a statistical fluke, given the aleatoric uncertainties, though it may also potentially be traced back to the larger Fisher information for $\sigma_8$ than for $\Omega_m$ at these additional $k\sim0.2-0.3h\Mpc^{-1}$ modes: while mode coupling induced by the additional low-k modes causes some modest information loss for $\Omega_m$, the information in the additional low-k modes themselves potentially (more than) compensates for that in the $\sigma_8$ case.

\begin{deluxetable*}{c|ccc|ccc|ccc}[htb!]
\tablecaption{RMSE values obtained from total matter density maps, relative to the network that is both trained and tested on no-monopole maps from SB28, which produces RMSE values of $0.04$ for $\Omega_m$ and $0.065$ for $\sigma_8$. The aleatoric uncertainties on these values are of order $10\%$.}
\label{tab:maps_RMSE}
\tablehead{
\multicolumn{1}{c|}{} & \multicolumn{3}{c|}{$\Omega_m$ no monopole} & \multicolumn{3}{c|}{$\sigma_8$ no monopole} & \multicolumn{3}{c}{$\Omega_m$ with monopole} \\
\cline{1-10}
\multicolumn{1}{c|}{\diagbox[width=3cm]{\textbf{Trained on}}{\textbf{Tested on}}} & \colhead{SB28} & \colhead{SB35 quadrants} & \multicolumn{1}{c|}{SB35} & \colhead{SB28} & \colhead{SB35 quadrants} & \multicolumn{1}{c|}{SB35} & \colhead{SB28} & \colhead{SB35 quadrants} & \colhead{SB35}
}
\startdata
SB28 & 1.0 & 1.18 & -   & 1.0  & 0.98  & -   & 0.19 & 0.21 & - \\
SB35 quadrants & 1.01 & 1.03 & -   & 1.04 & 0.96 & -   & 0.21 & 0.22 & - \\
SB35 & -   & -    & 0.81 & -    & -    & 0.73 & -   & -   & 0.13 \\
\enddata
\end{deluxetable*}

Further, we find that inference from full-field SB35 maps scales less strongly than the na\"ive expectation of the square root of the volume (in this case, area, as the slice thickness is kept fixed), namely, by a factor of two. Instead, as seen in Table \ref{tab:maps_RMSE}, the RMSE of full-field SB35 maps is lower merely by a factor of $\approx1.3$ than the RMSE of the SB35 quadrants (as well as that of the full-field SB28 maps). As discussed above, two types of explanations for this are possible. It may be the case that the constraining power can, in principle, be improved by the expected volume factor in some complicated high-dimensional manner, to which we are insensitive since we only constrain the marginals of the full posterior. It may, however, be the case that mode coupling, including between the additional low-k modes and the modes that are resolved in both map sizes, induces correlations between the four quadrants of each SB35 $(50\hMpc)^2$ map, such that their combined area does not contain four times as much information. Distinguishing these two possibilities requires more research, which we reserve for later work. Regardless, it is worth noting that the improvement between SB28 and SB35 is stronger than that which we observed with the power spectrum in \autoref{sec:Pk}. All in all, the RMSE values obtained for SB35, even with the no-monopole maps, are factors of $\approx1.5-3$ better than those with the power spectrum seen in \autoref{sec:Pk}. This accentuates the advantage of field-level inference in extracting more complete information than summary statistics such as the power spectrum.

We now move to discussing the results from unnormalized maps\footnote{Note that they are still normalized globally before being fed to the CNN, following best practices. However, since the normalization is global over all the maps, each map retains the monopole information. This is opposed to our no-monopole (`normalized') maps, which are first normalized individually to erase the monopole information.}, which contain the value of the monopole, namely the total mass of the slice from which the map is made. We focus on the in-distribution tests along the diagonal in the right section of Table \ref{tab:maps_RMSE}. We find that the monopole bears no additional constraining power for $\sigma_8$ and therefore omit quoting the exact RMSE values for this case in Table \ref{tab:maps_RMSE}, which are very similar to the ones quoted for the no-monopole $\sigma_8$ case. For $\Omega_m$, however, the monopole plays a special role, as both are direct measurements of total mass content. For the full simulation volume, the total mass content exactly tracks $\Omega_m$ by definition. While the correlation is not without scatter for our slices, which are sub-volumes of the full simulation volume, the constraining power of unnormalized maps is evidently much stronger than that of no-monopole maps in this case, with RMSE improvement factors of $\approx4-6$ for the in-distribution tests shown along the diagonal in Table \ref{tab:maps_RMSE}. In fact, a direct comparison of the relative RMSE values (RMSE normalized by the fiducial $\Omega_m=0.3$) along the diagonal -- 2.5\%, 3\% and 1.7\% -- and the corresponding fractional scatter of the monopole values -- 6.2\%, 8.4\% and 4.2\%  -- suggests that the raw RMSE values track the scatter of the monopole values while nevertheless being smaller than them. We interpret this to indicate that these networks use the monopole values as a baseline, allowing them to provide better predictions than the no-monopole case, but also apply a `correction' that uses information from the detailed structure of the maps, such that the eventual constraints are tighter than the monopole scatter itself.

We note that in this unnormalized case, there is some evidence that the results from SB35 quadrants are worse than those from full-field SB28 maps (0.22 versus 0.19 in Table \ref{tab:maps_RMSE}) in a more significant way compared to the no-monopole (normalized) case (1.03 versus 1.0), although this may well be a statistical fluke, given the aleatoric uncertainties. If true, however, this can be seen as a manifestation of mode coupling leading to larger cosmic variance among SB35 quadrants -- in particular, for the monopole itself. We also note that the scaling between the full-map SB35 RMSE value and that from the SB35 quadrants (a factor of $0.22/0.13\approx1.7$ improvement) is closer to ideal than in the no-monopole case ($1.03/0.81\approx1.3$), likely tracking the factor of $1.76$ lower monopole scatter that we find in the larger slices. Finally, we note that the non-diagonal entries, which represent cross tests between SB28 and quadrants of SB35, do not indicate significant out-of-distribution effects here either, similarly to the no-monopole case.

We close this section with a test of whether the results improve with more data. For SB28, we can double the training set by using all 2,048 simulations rather than only 1,024 of them, as done above. For SB35, we can double the training set by using 10 slices per dimension instead of merely 5 slices, as done above. As expected from the convergence study in \citet{BairagiA_25a}, we do observe improved constraints when the larger training sets are used. In the case of no-monopole $\Omega_m$, the RMSE drops by $\approx13\%$, in the case of no-monopole $\sigma_8$, the drop is only $\approx5\%$, and in the case of $\Omega_m$ with monopole, the drop is by $\approx22\%$. If we assume that the power-law scaling identified in \citet{BairagiA_25a} between posterior widths and the number of simulations holds in our case, these improvements correspond to power-law indices of $\approx0.2$, $\approx0.08$, and $\approx0.38$, respectively, which are not dissimilar from those found in \citet{BairagiA_25a}.

\subsection{Galaxy graphs}
\label{sec:graphs}
Recently, Graph Neural Networks (GNNs) have been proposed as an alternative method for constraining cosmological parameters from the spatial distribution of halos and galaxies \citep{CosmoGraphNet,deSantiN_23a, MIT_CosmoBench,CosmoBench,CosmoTDL}. GNNs are able to effectively capture the small-scale clustering information from the distribution of halos and galaxies, since, unlike CNNs, which are bound by the grid size, they do not impose a cutoff in the clustering scale. Moreover, GNNs have been shown to exhibit a fair degree of robustness for both halo and galaxy distributions with distinct halo finders or subgrid physics models \citep{Shao2023,deSantiN_23a}. Here, we compare the constraints on $\Omega_m$ and $\sigma_8$ using GNNs on galaxy graphs created from the SB28 and SB35 simulations.

To generate galaxy graphs that consist of galaxies with similar properties across the two simulation sets, we impose a global threshold and select subhalos that have at least 20 star particles, following \cite{CosmoGraphNet}, which corresponds to a mass cutoff of $\approx1.5\times10^8\hMsun\times\Omega_b/0.049$. This choice yields a similar median galaxy number density of $n_{\rm{gal}} \approx 3\times10^{-2}(\hMpc)^{-3}$ for both SB28 and SB35, resulting in a median galaxy number of $\approx500$ for SB28 and $\approx4000$ for SB35. We then generate a radius graph, connecting all galaxies from this sample that are within a certain linking radius $r_{\rm{link}}$ as neighbors. We encode the positional information of galaxies to the edge features via the lengths and angles of the edges. Specifically, given two galaxies with positions $\mathbf{x_i}$ and $\mathbf{x_j}$, and their displacement $\mathbf{d_{ij}} = \mathbf{x_i}-\mathbf{x_j}$, we select the following edge features $\{|\mathbf{d_{ij}}|/r_{\rm{link}}, \alpha_{ij}, \beta_{ij}\}$, where $\alpha_{ij} = \mathbf{x_i}\cdot\mathbf{x_j}/|\mathbf{x_i}||\mathbf{x_j}|$ and $\beta_{ij} = \mathbf{x_i}\cdot\mathbf{d_{ij}}/|\mathbf{x_i}||\mathbf{d_{ij}}|$ \citep{CosmoGraphNet,UniversalScalars}. Inspired by \cite{deSantiN_23a}, for the node features of the graph we use the z-direction velocity $v_z$ as a simple stand-in for the line-of-sight peculiar velocity in the far field.

The generated graphs are then fed to the GNNs implemented in \cite{SW_CosmographNet} and \cite{CosmoGraphNet}, which update the given node and edge features by passing messages between neighboring nodes. For training, we use the \citet{JeffreyN_20a} loss function and the exact same split of the data into training, validation, and test sets as in the preceding subsections. The batch size is fixed at 16 samples, and we train the neural network for 300 epochs with the ADAM optimizer \citep{Kingma2014}, adopting cyclic learning rates \citep{CyclicLR}. We also vary the following hyperparameters: linking radius cutoff $r_{\rm{link}} \in [0.005L_{\rm box}, 2.5\hMpc]$, where $L_{\rm box}$ is the box side length\footnote{The best value of $r_{\rm{link}}$ from hyperparameter optimization yields values of $1.25\hMpc$ and $2.15\hMpc$ for SB28 and SB35, respectively, which results in a median value of approximately 700 and 30,000 edges, respectively, a larger ratio than that of the number of nodes. We also find that these values are close to the upper bound of $2.5\hMpc$ for SB35. Although allowing a larger linking length may lead to better optimization, we note that the computational complexity increases significantly as the connectivity of graphs grows, and it may also lead to inefficiencies such as oversquashing of information \citep{oversquashing_curvature_2021}.}, learning rate $\in [10^{-8}, 10^{-4}]$, weight decay $\in [10^{-8}, 10^{-4}]$, number of layers $n_{\mathrm{layers}} \in [1, 5]$, and hidden channels $\in \{64, 128, 256\}$. These hyperparameters are tuned over 100 trials with \textsc{Optuna} \citep{Optuna}, and we show the results (for the test set) of the network with the best validation loss.

\begin{figure}[ht!]
\includegraphics[width=0.46\textwidth]{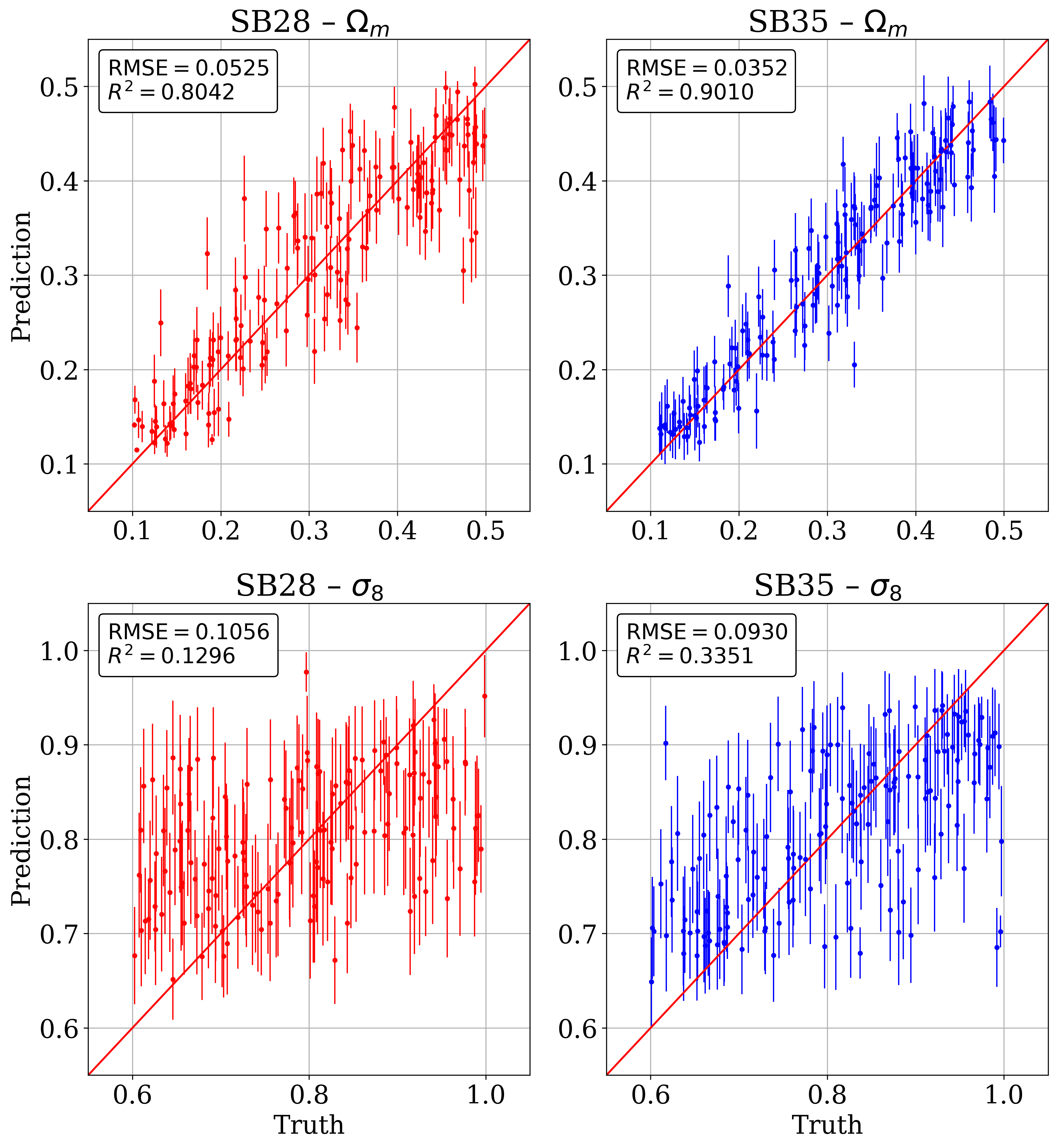}
\caption{Inference results on the cosmological parameters $\Omega_m$ (top) and $\sigma_8$ (bottom) from galaxy graphs using GNNs. We observe an increased constraining power with SB35 (right) compared to SB28 (left), but the scaling is significantly weaker than the square root of the volume ratio, as discussed in \autoref{sec:graphs}.}
\label{fig:inference_GNN}
\end{figure}

We present the results from this exercise in \autoref{fig:inference_GNN}, which has a similar format to \autoref{fig:inference_Pk}, with SB28 results on the left, SB35 results on the right, $\Omega_m$ at the top, and $\sigma_8$ at the bottom.
With SB35, we obtain an RMSE for $\Omega_m$ of $0.035$, which compares favorably to $0.095$ from the power spectrum but is slightly inferior to the RMSE of $0.029$ obtained from the (no-monopole) density maps, which are discussed in \autoref{sec:Pk} and \autoref{sec:projected_maps}, respectively. For $\sigma_8$, we obtain an RMSE of $0.093$, which indicates lower constraining power compared to the values of $0.075$ from the power spectrum and $0.052$ from the (no-monopole) density maps.

As seen in these previous sections, we find an improvement in the constraining power between SB28 and SB35; the RMSE for $\Omega_m$ improves by a factor of $1.4$, and while the improvement for $\sigma_8$ is more modest, graphs from SB35 clearly have some constraining power (with a coefficient of determination $R^2=0.34$), which those from SB28 lack almost entirely. These RMSE improvements, however, are significantly smaller than the na\"ive scaling with the square root of the volume ratio, $\sqrt{8}\approx2.8$.

We interpret both the weaker scaling and the larger RMSE values compared to the experiments in the previous sections (for the most part, with the exception of $\Omega_m$ compared to the power spectrum) as stemming from the stronger dependence of the galaxy graphs on astrophysical parameters. The reason is that, as opposed to the matter power spectrum and the projected mass maps, which, while certainly sensitive to baryonic feedback \citep{DelgadoA_23a,GebhardtM_24a,GebhardtM_26a}, are based on the total matter distribution (which is dominated by dark matter), the graphs are directly based solely on galaxy properties. While galaxy positions and velocities may be tightly related to the gravitationally dominant dark matter component, the way they populate dark matter halos depends heavily not only on the cosmological parameters but also on the subgrid physics parameters. In particular, for a fixed combination of cosmological parameters, the resulting galaxy samples used in this exercise, namely galaxies with at least $20$ stellar particles, may differ significantly from each other in both numbers and environments across different choices of astrophysical parameters. This reduces the constraining power on individual cosmological parameters such as $\Omega_m$ and $\sigma_8$ through degeneracies between the parameters with respect to their effects on the resulting graphs. These degeneracies are also the likeliest cause for the weaker scaling between the $25\hMpc$ and the $50\hMpc$ volumes than that found in the previous sections (see the discussion in \autoref{sec:theory}).

Fundamentally, however, these limitations on our inference exercise originate in a significant advantage of this approach: its reliance on luminous galaxies rather than total (including dark) matter makes it much closer to the directly observable realm (even though we have not employed any detailed modeling to align the graphs with the properties of actual galaxy surveys). A distinct possibility that might also be contributing to the weaker volume scaling is the optimization of the GNNs, which is known to be harder to achieve for larger graphs. In our case, the eightfold larger number of nodes in the SB35 graphs, and the even larger ($50$-fold) number of edges due to the larger physical linking length, may lead to less optimized GNNs than those based on SB28 due to significantly higher computational and memory demands, as well as the effectively enlarged receptive field per node. These factors are known to complicate training and convergence on large-scale graphs \citep{gnn_Duan_2022, gnn_guan22d}. Another possible contribution to the weaker scaling is the fact that our GNNs extract information mostly from small scales and are limited in capturing long-range interactions \citep{gnn_LRI_Li_2023, gnn_LRI_benchmark}, in combination with enhanced mode coupling in SB35 compared to SB28, which reduces the number of independent small-scale modes. We finally comment that we repeated the training of the SB28 graphs utilizing the full sample of 2,048 SB28 simulations for comparison, and obtained $\approx5\%$ improvements on the RMSE values relative to the results shown in the left column of \autoref{fig:inference_GNN}. This indicates that there is potential for further improvements with larger future simulation sets, as well as with new methodologies tailored to extract more information from the graph \citep{CosmoTDL}.

\subsection{Thermodynamical properties of massive halos}
\label{sec:massive_halos}

Here we change gears and move from constraints from the global distribution of matter and galaxies in the simulation volumes to an investigation of thermodynamical properties of individual massive halos ($M_{200c}>10^{13}\hMsun$)\footnote{Where $M_{200c}$ is the mass enclosed by a sphere with an overdensity of $200$ with respect to the critical density, centered on the most bound particle in the halo.}. In this case, we extend the inference to the marginals of all 35 (or 28) parameters that are varied in the simulations, rather than focusing on $\Omega_m$ and $\sigma_8$ alone. However, we begin with an emulation task and follow it with an inference task that relies on the developed emulator.

One advantage of the SB35 set over SB28 for studies based on massive halos is that the evolution of these halos is more physically accurate in SB35 due to the lack of low-k modes in SB28, as suggested by \autoref{fig:c_vs_Mvir}. In particular, we might expect that the nonphysical offset in the concentration--mass relation seen in \autoref{fig:c_vs_Mvir} translates to nonphysical offsets in the thermodynamical properties of the halos, such as their Compton $Y_{\rm SZ}$ from the Sunyaev-Zeldovich effect \citep{WadekarD_23a}.

\begin{figure}[ht!]
\gridline{\fig{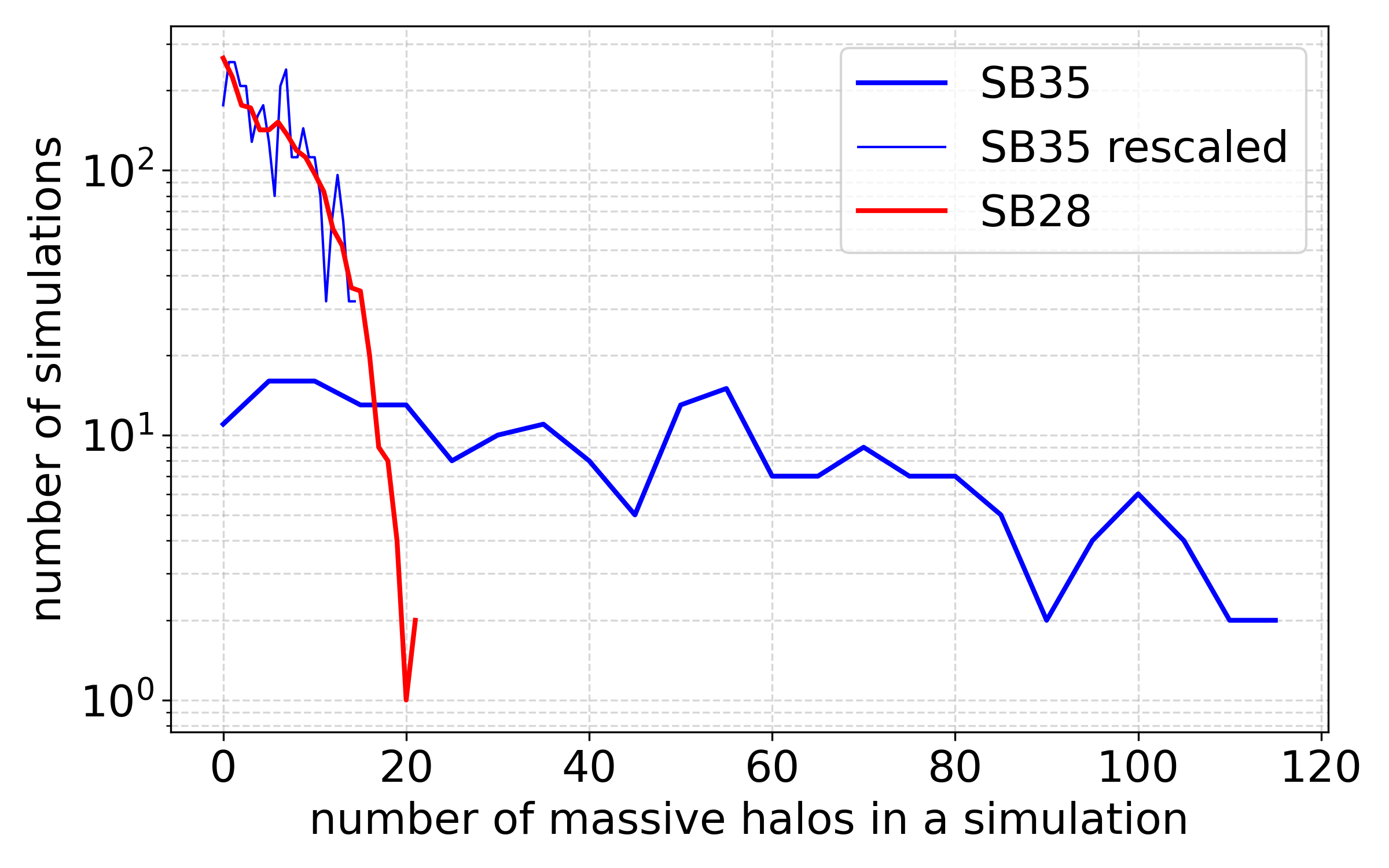}{0.48\textwidth}{(a)}}
\gridline{\fig{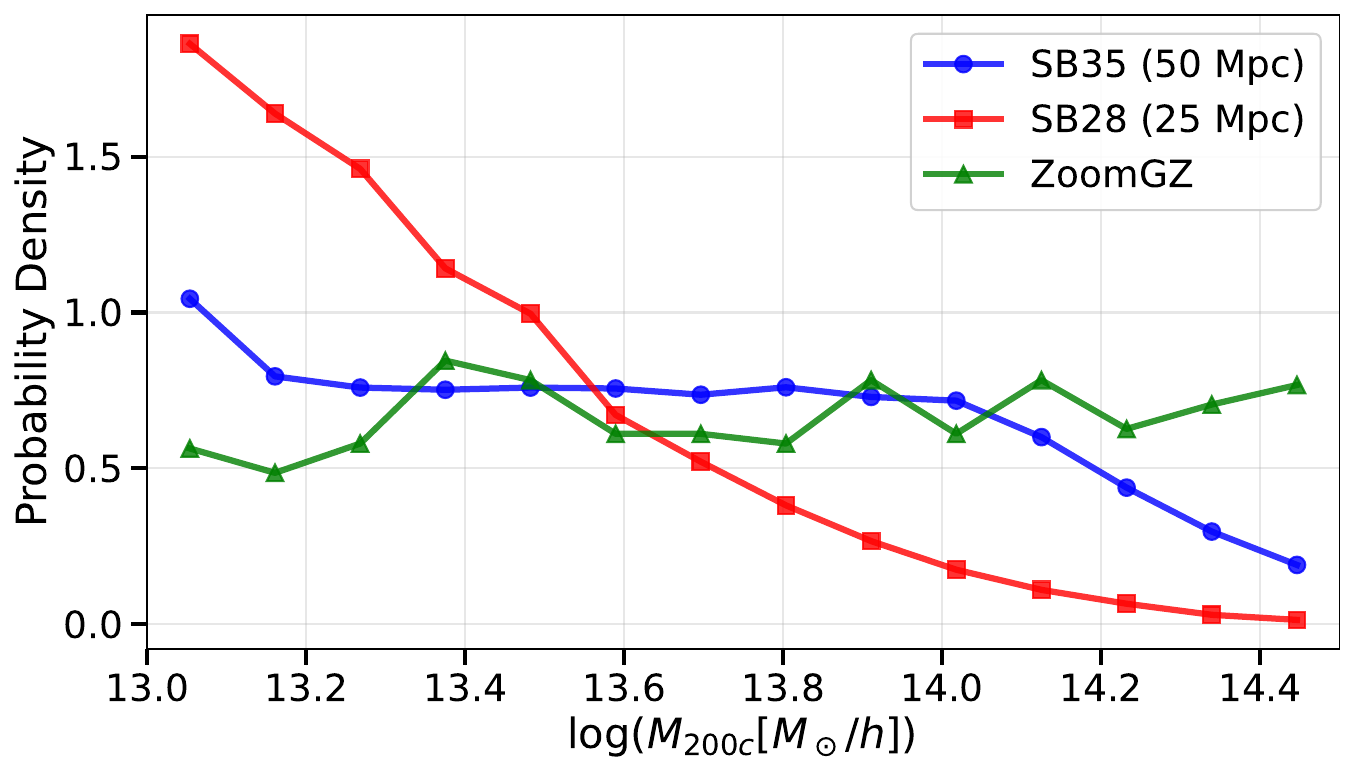}{0.46\textwidth}{(b)}}
\caption{{\it Top:} A comparison of the distribution of the halo count per simulation box of halos with $M_{200c}>10^{13}\hMsun$, in the $25\hMpc$ SB28 simulations (red) and $50\hMpc$ SB35 simulations (blue) at $z=0$. The larger volumes yield a significantly higher count of massive halos, but when rescaled appropriately by the volume and the total number of simulations in each set (thin blue), the distribution of halo counts is consistent with that of the smaller volumes. {\it Bottom:} Normalized distributions (probability density-like) of the masses of the halos used to train the Gaussian Process emulators for each of the SB sets as well as for the \citet{LeeM_24a} zoom-in simulations (green). For SB28 we use almost all of the halos with $M_{200c}>10^{13}\hMsun$ in the simulation set, hence the mass distribution is very skewed, mirroring the generic halo mass function, while for SB35 we are able to select nearly 10,000 halos while sampling almost uniformly (in logarithmic space) from the full mass range.}
\label{fig:Nhalos}
\end{figure}

Another advantage of SB35 over SB28 is the sheer number of halos. \autoref{fig:Nhalos}(a) presents distributions of the number of such halos among our different simulation boxes for each of the SB28 (red) and SB35 (blue) sets and demonstrates that the number of halos per simulation is, as expected, about eight times larger in the $50\hMpc$ boxes than in the $25\hMpc$ boxes. Since the SB28 set comprises twice as many simulations as SB35 does, the total number of massive halos in the new SB35 set is just about four times larger than the total in the SB28 set -- 41,975 versus 11,258 massive halos. For technical reasons\footnote{Gaussian Process regression is limited by computationally expensive matrix inversions that scale as $N_{\rm samples}^3$, reducing the total number of halos we can use in our training to $\approx10^4$.}, however, in the following analysis we use a similar number of halos from each of these two sets, roughly $10^4$ halos. To achieve this number of halos from SB28, we must keep almost all of the halos, resulting in a halo mass distribution that is seen in \autoref{fig:Nhalos}(b) (red), which is naturally skewed towards lower masses due to the well-known shape of the halo mass function. However, from SB35 (blue), we exclude some of the $M_{200c}\lesssim10^{14}\hMsun$ halos in order to achieve a nearly uniform (in logarithmic space) sampling of halo masses in that range while keeping all of the halos with $M_{200c}\gtrsim10^{14}\hMsun$. For a fixed total number of halos from the two sets, $\approx10^4$, this more uniform mass distribution achievable with SB35 is advantageous, as discussed below.

In this section, we have an additional important reference point from previous work, with which we compare the new SB35 simulations: the CAMELS-ZoomGZ zoom-in simulation suite \citep{LeeM_24a}, which focuses specifically on massive halos and employs CARPoolGP, a novel technique to mitigate sample variance in a large model parameter space. CAMELS-ZoomGZ comprises merely 768 simulations of individual halos (targeting the mass range $10^{13}\hMsun<M_{200c}<10^{14.5}\hMsun$); however, thanks to their unique design (specially designed correlations between halos) and CARPoolGP, their statistical power is much higher than that of a random selection of halos, as we demonstrate below.

To create an emulator of the Sunyaev-Zeldovich Y-M relation, we train a Gaussian Process based on each of our SB28 and SB35 halo samples and evaluate these emulators at the fiducial set of parameters of the IllustrisTNG model, which was not seen during the training process.
The calculation of the target quantity for the emulator, the integrated $Y_{\rm SZ}$, and the training of the Gaussian Process are performed as in \citet{LeeM_24a}.
We compare these emulations to the results from the emulator trained in \citet{LeeM_24a} based on the CAMELS-ZoomGZ halos (presented in their Figure 8), as well as to the raw (binned) results from the TNG300 simulation, which provides the best estimate for the `ground truth' of this relation in the fiducial IllustrisTNG model, since it has a very similar resolution to the CAMELS simulations but a much larger volume and hence halo count.

\begin{figure*}[ht!]
\includegraphics[width=0.99\textwidth]{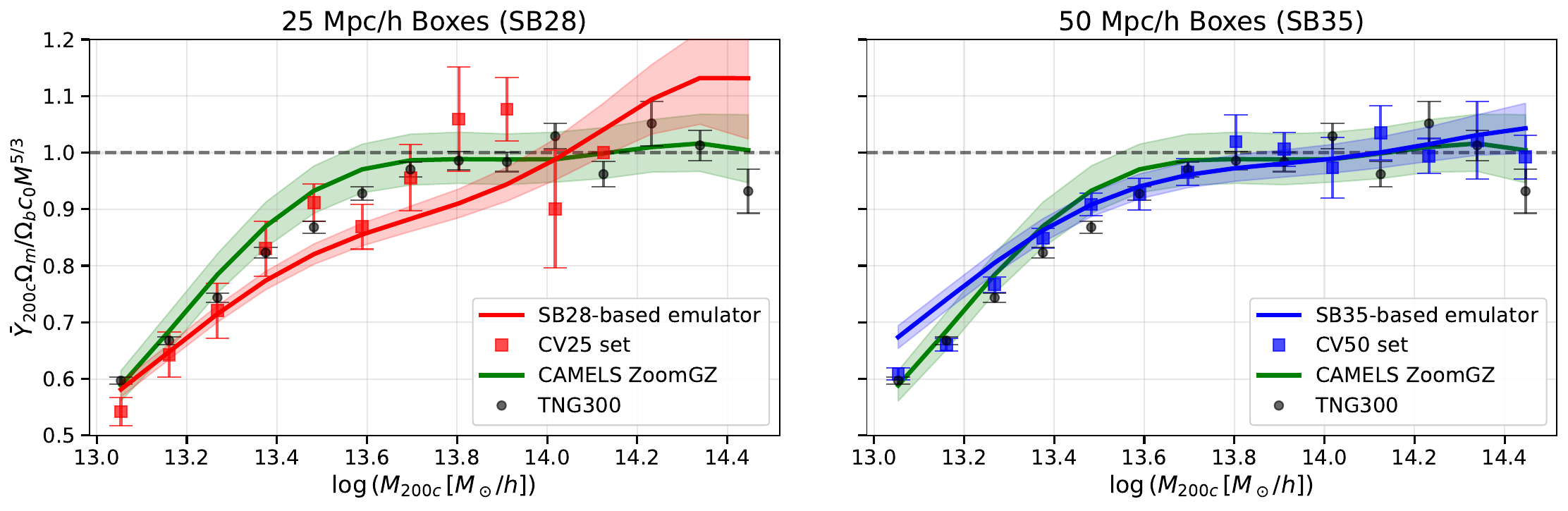}
\caption{A comparison of thermal Sunyaev–Zeldovich $Y$–$M$ relations at $z=0$ for the fiducial IllustrisTNG model, as inferred from different simulations. Both panels include binned actual measurements from the TNG300 simulation (black) and from our respective CV sets (red and blue squares), which serve as `ground truth', as well as the relation emulated with CARPoolGP based on the CAMELS-ZoomGZ zoom-in simulations (green; \citealp{LeeM_24a}). To these baselines, we add the emulated relation based on halos in the SB28 set (left, red) and based on halos in the SB35 set (right, blue).}
\label{fig:YM_relation}
\end{figure*}

These results are presented in \autoref{fig:YM_relation}, where the left panel shows the SB28 results (red) and the right panel shows the SB35 results (blue). In each of these, we present the binned results from the corresponding CV sets (which contain a total of 121 and 888 halos in this mass range for the $25\hMpc$ and $50\hMpc$ boxes, respectively), which serve as versions of the `ground truth' for the fiducial model (squares with error bars), as well as the emulation result (solid) and its predictive uncertainty (shaded). In both panels, the results are compared with the CAMELS-ZoomGZ emulation (green) and TNG300 (black).

Our findings based on \autoref{fig:YM_relation} are as follows. First, we note that both CV sets appear statistically consistent with the TNG300 `ground truth', and that, as shown in \citet{LeeM_24a}, the CAMELS-ZoomGZ emulation successfully reproduces that `ground truth'. Second, the SB35 emulation also closely matches that same relation, with only the lowest-mass bins being $\approx3-4\sigma$ too high. In contrast, the SB28 emulation, which uses an almost identical total number of halos as SB35, results in a significantly inferior match to the actual simulation results. It is notable that the SB28 emulation performs best at low masses; this is because SB28 has more significant constraining power in the mass range where there are numerous halos\footnote{We confirm this hypothesis by training a separate GP emulator on a sample from SB35 that mimics the SB28 mass function at $M_{200c}<10^{14}\hMsun$. This results in a significantly better match to TNG300 where the deviations are seen in \autoref{fig:YM_relation}, $M_{200c}\lesssim10^{13.3}\hMsun$ (for brevity, not shown explicitly).}. However, since the halo number drops significantly with increasing halo mass (\autoref{fig:Nhalos}(b)), the achievable accuracy with SB28 decreases too. Third, we compare the predictive uncertainties of the emulators and note that they are indeed very small with SB28 in the low-mass end, but grow with mass. More significantly, we find that the predictive uncertainty with SB35 is much more uniform and is comparable to, but even smaller than, those from CAMELS-ZoomGZ. On one hand, since the CAMELS-ZoomGZ simulations were created and designed specifically for maximizing the return on the Y-M relation, this is a feat for the SB35 set that is achieved by the sheer combined volumes ($1024\times(50\hMpc)^3\approx(504\hMpc)^3$) and the number of halos it contains. In fact, since we only use $\approx1/4$ of the total number of SB35 halos here, we can expect the uncertainties (and correspondingly deviations from the `ground truth') to be potentially roughly halved with the full sample. On the other hand, this demonstrates the power of CAMELS-ZoomGZ and CARPoolGP, which achieve comparable uncertainties with only 768 halos, thanks to their specialized design.

Next, we use the SB35 massive halos for parameter inference based on their emulated thermodynamical radial profiles, following \citet{Hernandez-MartinezE_25a}. To this end, we measure five spherically-averaged quantities, centered around the gas density peak, in 40 radial bins\footnote{We note that for the 2D quantities, i.e., X-ray surface brightness and Compton-y, we used the CGM Toolkit \citep[][\url{https://github.com/ethlau/cgm_toolkit}]{LauE_25a,LauE_25b,LauE_25c}. This package computes quantities in 3D spherical shells defined by radial bins, then applies interpolation and an Abel projection to obtain 2D profiles. When projecting a spherical 3D profile into 2D using the Abel transform, the outermost radial bin corresponds to a shell of zero thickness, and thus its projected value is always zero. As a result, these quantities effectively have 39 bins instead of 40.} that are logarithmically spaced in the range $0.01R_{200c}$ to $2.5R_{200c}$ for each of the 9332 halos in the SB35 sample discussed above. The five quantities are gas temperature, density, and metallicity, as well as Compton $y_{\rm SZ}$ and X-ray surface brightness. We then train a Gaussian Process for each combination of quantity and radial bin, namely $3\times40+2\times39$ emulators, as a function of the 35 simulation parameters as well as the halo mass. After training, we generate $3\times10^4$ emulated profiles at a fixed $M_{200c}=10^{14}\hMsun$ while randomly sampling the 35-dimensional parameter space. We then train neural networks (following the same methods as in \citealp{Hernandez-MartinezE_25a}) to infer the simulation parameters from the combination of these five types of profiles.

The focus of this analysis is the comparison of the constraining power across different radial ranges. \citet{Hernandez-MartinezE_25a} found that the constraining power is concentrated towards the inner parts of the halos, such that even the exclusion of the information within just $0.1R_{200c}$ leads to appreciably smaller correlation coefficients, albeit more so for some parameters than others. Here, SB35 allows us to extend the radial range well beyond the virial radius, which is not possible with the zoom-in simulations of CAMELS-ZoomGZ, as they are designed to resolve the halos inside their virial radius but quickly become `contaminated' by low-resolution particles outside of it. Hence, here we use SB35 to explore the constraining power of extended profiles on simulation parameters. To this end, we split the emulated profiles into an inner and an outer part at the bin corresponding to $0.9 \times R_{200}$. This ensures that the input vector for both the interior and outskirts regions has a dimensionality larger than the number of parameters to be inferred, which is required for the network to successfully recover them. In practice, this yields an input size of 160 for the interior and 38 for the outskirts.

\begin{figure*}[ht!]
\includegraphics[width=0.99\textwidth]{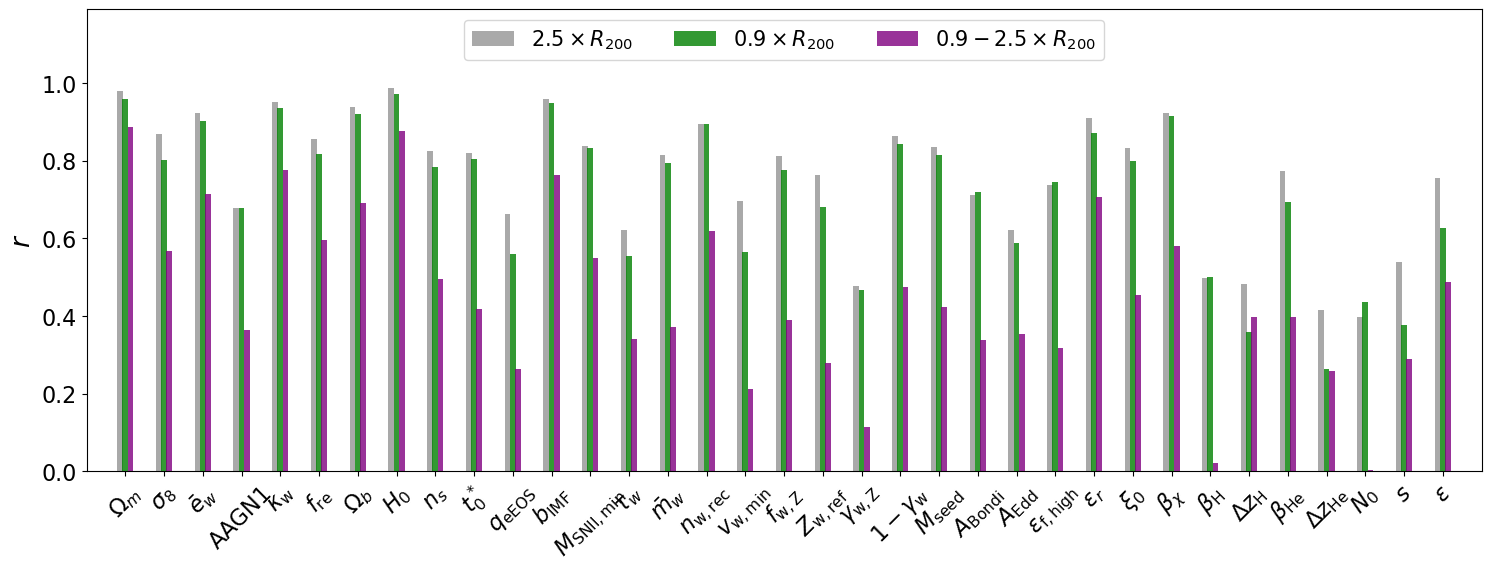}
\caption{Correlation coefficients between predicted and true values of 35 parameters, based on inference networks that are trained on halo thermodynamical radial profiles. Three distinct networks use three radial ranges: the full range up to $2.5R_{200c}$ (grey), the inner range up to $0.9R_{200c}$ (green), and the outer range between $0.9R_{200c}$ and $2.5R_{200c}$ (purple). While adding the outer profiles to the inner ones adds little constraining power, the outer profiles on their own do contain a significant amount of information for constraining many of the parameters, even if not as strongly as the inner profiles.}
\label{fig:inference_profiles}
\end{figure*}

We present the results in \autoref{fig:inference_profiles} by means of the Pearson correlation coefficient between predicted and true parameter values on the test set, analogously to Figures 6 and 7 in \citet{Hernandez-MartinezE_25a}. We first note that the results from what we define here as the `inner' profiles, up to $0.9R_{200c}$, are similar to those from \citet{Hernandez-MartinezE_25a} based on the CAMELS-ZoomGZ emulator. To test the possible role that the 7 additional parameters that are varied in SB35 but not in CAMELS-ZoomGZ might play in the slightly worse performance found here (typically $\Delta r\approx0.05$), we perform a benchmark inference based on emulated profiles (from the SB35-based emulator) that only sample the same 28-dimensional parameter space that is common with CAMELS-ZoomGZ. We indeed find (but do not show explicitly here) that the performance difference is partially recovered in this experiment, indicating the existence of some degeneracies between the new 7 parameters and the common 28 parameters in terms of their effects on halo profiles.

\autoref{fig:inference_profiles} also shows that, in line with \citet{Hernandez-MartinezE_25a}, limiting the data to just the outer parts ($0.9R_{200c}$ to $2.5R_{200c}$) results in sometimes significantly lower correlation coefficients. However, significant constraining power remains, in particular for some parameters, such as $\Omega_m$, $H_0$, and the speed of galactic winds $\kappa_{\rm w}$. That the beyond-$R_{200c}$ regime contains useful information for parameter constraints is important for future applications of this approach to observational data, since resolving a large number of halos down to their innermost parts (e.g.~$0.01R_{200c}$) will be impractical for the foreseeable future with either X-ray facilities (probing gas density, temperature, and metallicity) or in the millimeter (probing gas density and temperature through the SZ effect).

\section{Summary and Conclusions}
\label{sec:summary}

\begin{figure}[ht!]
\includegraphics[width=0.46\textwidth]{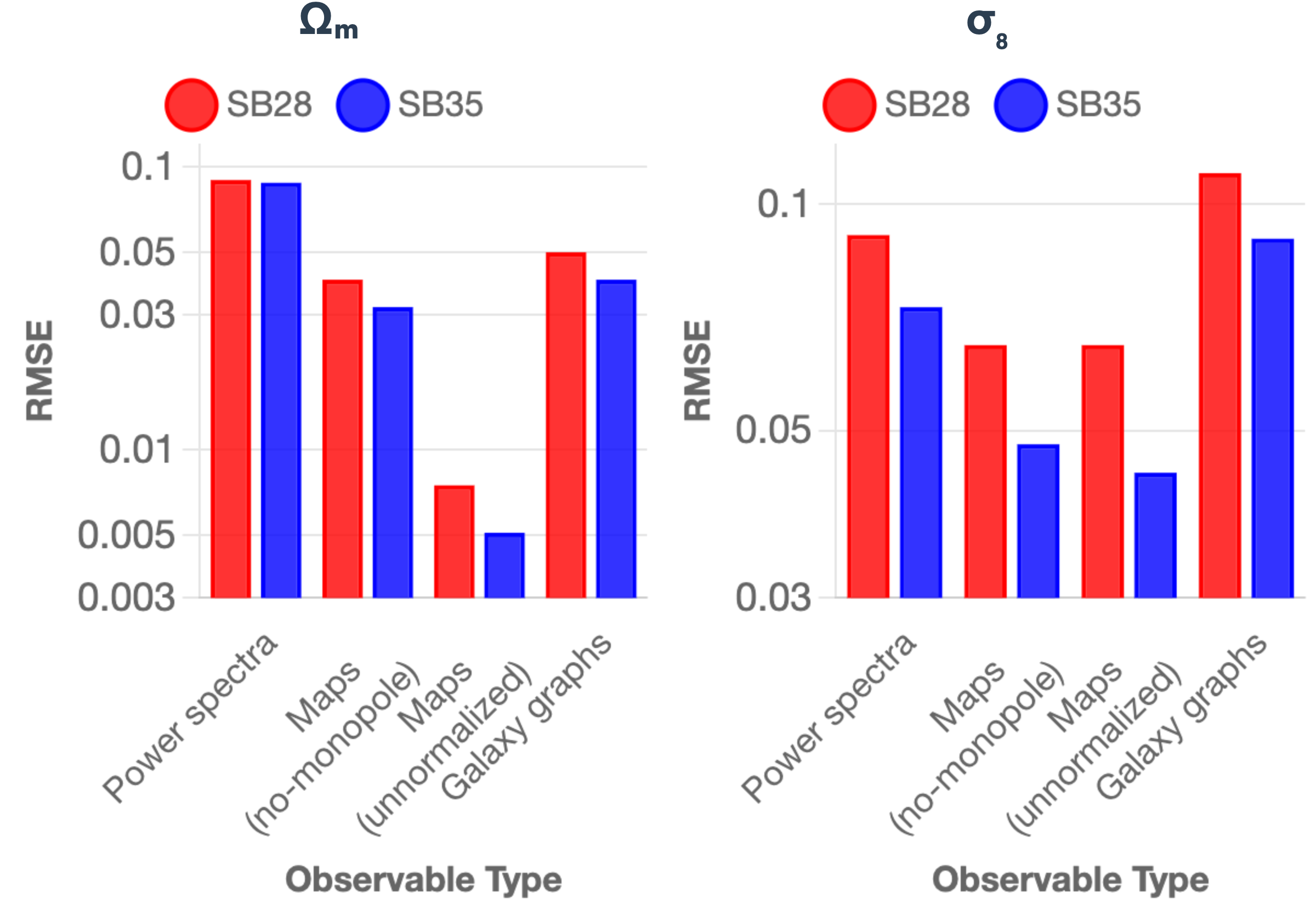}
\caption{A visual representation of the data in Table~\ref{tab:summary_RMSE}: summary of RMSE values of parameter inference based on various data. The relative constraining power of the different observables differs between $\Omega_m$ (left) and $\sigma_8$ (right). Each pair of red (SB28) and blue (SB35) columns corresponds to a given type of input data, showing that the improvement with increasing volume is weakest with the power spectra and strongest with the unnormalized matter density maps.}
\label{fig:summary_RMSE}
\end{figure}

We have presented the new SB35 suite of the CAMELS project, designed to explore small-scale structure formation across a broad cosmological and astrophysical parameter space, with increased volume ($(50\hMpc)^3$) and parameter space (35 parameters; Table \ref{tab:parameters}) relative to previous CAMELS simulations ($(25\hMpc)^3$; up to 28 parameters). Our main findings from comparisons between these new simulations and previous CAMELS simulations (Table \ref{tab:simulations}) can be summarized as follows:

\begin{itemize}
\item \textbf{Simulation overview and consistency:} Global properties of the SB35 suite, including the matter power spectrum, star formation rate density, and the stellar mass function, are broadly consistent with the previous SB28 simulations where the coverage overlaps, while extending the coverage to higher mass halos and larger scales (\autoref{fig:global_properties}).
\item \textbf{Galaxy scaling relations:} The SB35 simulations show general agreement with SB28 on key galaxy scaling relations (stellar mass, size, metallicity, SFR) in the overlapping regime. The LH simulations, with reduced parameter variation, exhibit narrower scatter and some systematic offsets in the medians (\autoref{fig:galaxy_scaling_relations}).
\item \textbf{Thermal state of the IGM:} Variations in the ionizing background parameters produce measurable differences in the temperature--density relation of the IGM (\autoref{fig:T0_gamma}).
\item \textbf{Dark matter halos:} We quantify for the first time the bias in the concentration--mass relation in boxes of $(25\hMpc)^3$ (\autoref{fig:c_vs_Mvir}), which originates from missing large-scale modes and suppressed non-linear evolution. Further, the larger $(50\hMpc)^3$ volume in SB35 yields improved halo number statistics, particularly at the high-mass end (\autoref{fig:Nhalos}).
\item \textbf{Parameter inference from the matter power spectrum:} Constraints on $\Omega_m$ from the $z=0$ non-linear matter power spectrum barely tighten between SB28 and SB35, while those on $\sigma_8$ improve by a factor of $\approx1.25$, far short of the na\"ive $\sqrt{8}\approx2.8$ scaling with volume (\autoref{fig:inference_Pk}). The bulk of this modest gain comes from reduced sample variance in the $k$-bins common to both volumes rather than from the additional large-scale modes uniquely available in SB35. We trace the weak scaling primarily to strong mode coupling, which leaves the cosmic variance of the power spectrum ratio only $\approx15\%$ smaller in CV50 than in CV25 (\autoref{fig:CV}).
\item \textbf{Parameter inference from field-level maps:} Field-level inference from projected total matter density maps yields tighter constraints than the power spectrum, with the gain from SB28 to SB35 reaching factors of $\approx1.3$ for no-monopole maps and $\approx1.5$ when the monopole is retained. For $\Omega_m$, unnormalized maps are particularly powerful since the monopole directly tracks the total mass, though the network refines the prediction beyond what the monopole scatter alone would allow. Cross-tests between full-field SB28 maps and SB35 quadrants reveal no significant out-of-distribution biases (Table \ref{tab:maps_RMSE}, \autoref{fig:maps}).
\item \textbf{Parameter inference from galaxy graphs:} GNNs applied to galaxy graphs deliver a root-mean-squared error (RMSE) improvement of $\approx1.4$ for $\Omega_m$ and $\approx1.2$ for $\sigma_8$ between SB28 and SB35, which, again, falls well short of the na\"ive volume scaling. The constraining power is weaker than that of the matter-based inputs, likely because galaxy populations depend strongly on subgrid physics, introducing degeneracies between cosmological and astrophysical parameters. Optimization challenges associated with the much larger SB35 graphs (a factor of $\approx50$ more edges) may also contribute to the weaker-than-na\"ive scaling (\autoref{fig:inference_GNN}).
\item \textbf{Summary of parameter inferences:} For the convenience of the reader, in Table \ref{tab:summary_RMSE} and \autoref{fig:summary_RMSE} we provide a summary of the RMSE values from our various inference experiments, comparing both across the different observables and neural networks, as well as between the smaller volume SB28 suite and the newer, larger one, SB35.
\item \textbf{Thermodynamical properties of massive halos:} Gaussian Process emulators trained on SB35 halos accurately recover the $Y_{\rm SZ}–M_{200c}$ relation from TNG300 across most of the $10^{13}\hMsun<M_{200c}<10^{14.5}\hMsun$ mass range, while emulators trained on the same total number of SB28 halos perform significantly worse, a consequence of the much better sampling at high masses enabled by the larger volumes. Inference from thermodynamical radial profiles confirms that the inner regions of halos carry the bulk of the constraining power, but the outer regions (0.9–2.5 $R_{200c}$) retain substantial information for several parameters, including $\Omega_m$, $H_0$, and the galactic wind speed $\kappa_w$, which is an encouraging result for applications to observations that cannot resolve halo cores (\autoref{fig:YM_relation}, \autoref{fig:inference_profiles}).
\item \textbf{One-parameter variation diagnostics:} The 1P set provides guidance on how individual simulation parameters affect power spectra, SFRD, and galaxy--halo relations, which can guide the interpretation of trends in larger parameter grids (\autoref{fig:1P_Pk}, \autoref{fig:1P_SFRD}, \autoref{fig:1P_MsMh}).
\end{itemize}

\begin{deluxetable*}{c|ccc|ccc}[htb!]
\tablecaption{Summary of RMSE values of parameter inference based on various data. The percentage of improvement indicates the drop of the RMSE from inference based on the $(25\hMpc)^3$ boxes of SB28 to those of SB35 with $(50\hMpc)^3$.}
\label{tab:summary_RMSE}
\tablehead{
\colhead{} & \multicolumn{3}{|c|}{$\Omega_m$} & \multicolumn{3}{c}{$\sigma_8$} \\
\cline{2-7}
\colhead{} & \multicolumn{1}{|c}{SB28} & \colhead{SB35} & \colhead{\% improvement} & \multicolumn{1}{|c}{SB28} & \colhead{SB35} & \colhead{\% improvement}
}
\startdata
Matter power spectra (\autoref{sec:Pk}) & 0.09 & 0.088 & 2 & 0.091 & 0.073 & 19\\
Matter density maps, no-monopole (\autoref{sec:projected_maps}) & 0.04 & 0.032 & 19 & 0.065 & 0.048 & 27 \\
Matter density maps, unnormalized (\autoref{sec:projected_maps}) & 0.0075 & 0.0051 & 31 & 0.065 & 0.044 & 33 \\
Galaxy graphs (\autoref{sec:graphs})& 0.05 & 0.04 & 20 & 0.11 & 0.09 & 18 \\
\enddata
\end{deluxetable*}

The SB35 set significantly enhances the ability to support small-scale inference across a wider range of cosmic structures. It is made publicly available through several platforms (Uniform Resource Locator for direct download, Globus\footnote{\url{https://www.globus.org/}}, BinderHub\footnote{\url{https://binderhub.readthedocs.io/en/latest/}}) at \url{https://camels.readthedocs.io} to support further applications of machine learning and simulation-based inference in cosmology and galaxy formation.

\begin{acknowledgments}
We thank Adrian Bayer, Matt Ho, and Gus Beane for helpful discussions.
We are grateful to the Flatiron Institute's Scientific Computing Core for their support of the clusters on which these simulations were run.
The Flatiron Institute is supported by the Simons Foundation.
This work benefited from the ``Learning the Universe" collaboration, supported by the Simons Foundation.
SG and YJ acknowledge support from the Simons Foundation through grant PD-CMB-00011009.
BB acknowledges support from NSF grant AST-2009679 and NASA grant No.~80NSSC20K0500. This research was also supported in part by the National Science Foundation under Grant No. NSF PHY-1748958. BB is grateful for the generous support from the David and Lucile Packard Foundation and the Alfred P. Sloan Foundation.
DAA acknowledges support from the NSF CAREER award AST-2442788, an Alfred P. Sloan Research Fellowship, and the Cottrell Scholar Award CS-CSA-2023-028 by the Research Corporation for Science Advancement.
KN acknowledges support from JSPS KAKENHI grant 20H00180, 24H00002, 24H00241, JP25K01032, and the JSPS International Leading Research (ILR) project, JP22K21349. 
KN also acknowledges support from the Kavli IPMU, the World Premier Research Center Initiative (WPI), UTIAS, and the University of Tokyo.  
ChatGPT-5 \citep{ChatGPT20250815} and Claude-Opus-4.7 \citep{Claude20260516} were used to help draft the introduction, summary, and figure captions, as well as to format certain tables and figures and polish the language. All text generated by these tools was carefully reviewed, edited, and verified for accuracy by the authors, who assume full responsibility for the outcome.
\end{acknowledgments}

\begin{contribution}
SG ran the hydrodynamical simulations, guided the analysis, and wrote the manuscript.
YJ performed the analysis presented in \autoref{sec:projected_maps}.
BKO performed the analysis presented in \autoref{sec:global_properties}, \autoref{sec:scaling_relations} and \autoref{sec:Pk}.
MTT performed the analysis presented in, and co-wrote, \autoref{sec:radiation}.
MEL and EHM performed the analysis presented in \autoref{sec:massive_halos}.
JYL performed the analysis presented in, and co-wrote, \autoref{sec:graphs}.
FVN ran the N-body simulations and generated ancillary data.
BB, CL, DAA, FVN, KN and XS contributed to the analysis and interpretation of the results.

\end{contribution}

\appendix

\section{Single-parameter dependencies of various summary statistics}
\label{sec:dependencies}

\begin{figure*}[ht!]
\includegraphics[width=0.99\textwidth]{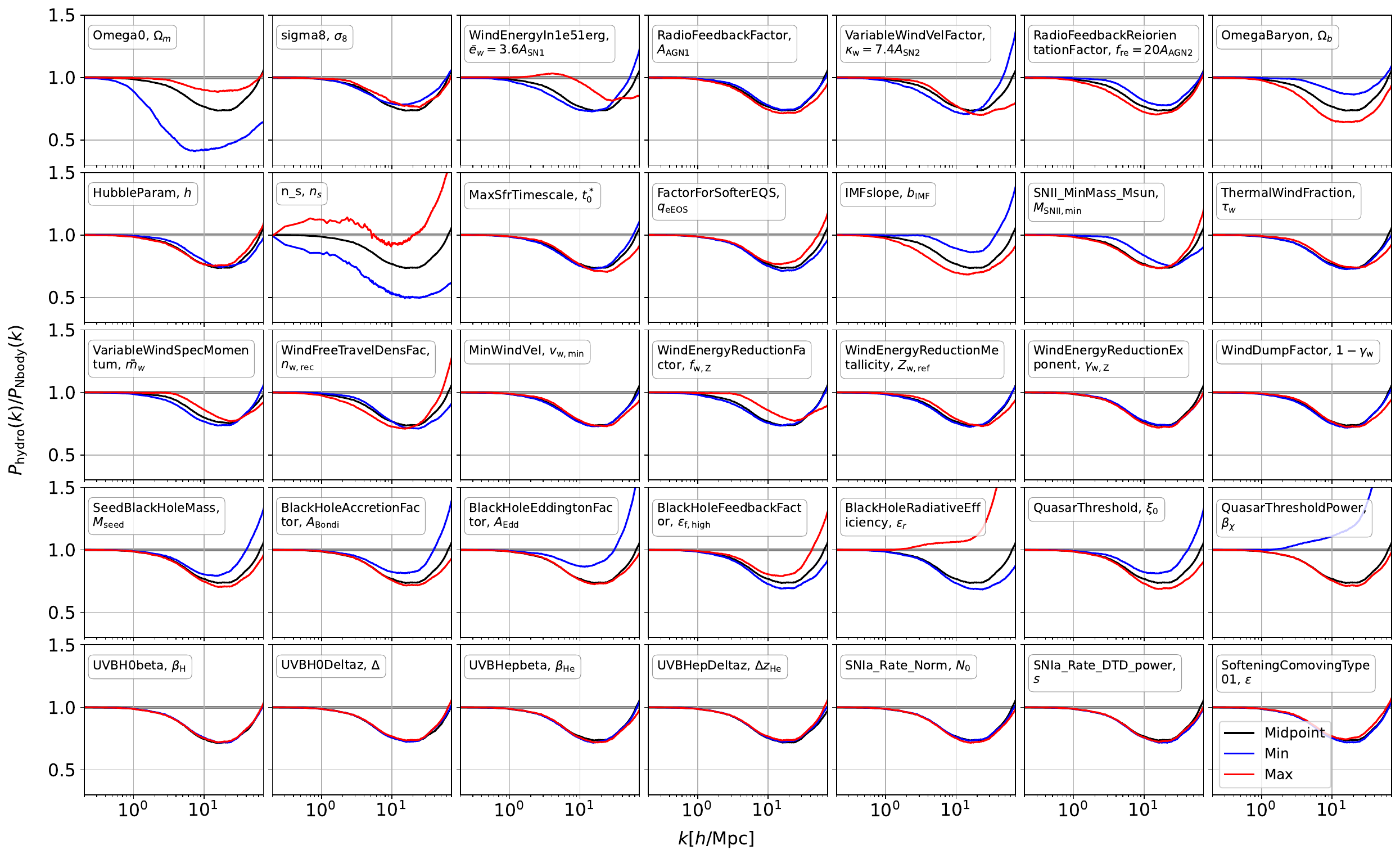}
\caption{Effect of single-parameter variations in the $50\hMpc$ 1P simulation set on the matter power spectrum at $z=0$. Each curve shows the ratio of the power spectra of a hydrodynamical simulation and its corresponding N-body simulation, where each parameter is varied separately in each panel. The simulation with the middle value for each parameter is shown in black, while blue and red curves denote the minimum and maximum parameter values that we use, respectively. Variations in both cosmological and astrophysical parameters, in particular AGN feedback, induce scale-dependent changes in the suppression of the power spectrum at $k \gtrsim 1h/\Mpc$.}
\label{fig:1P_Pk}
\end{figure*}

\begin{figure*}[ht!]
\includegraphics[width=0.99\textwidth]{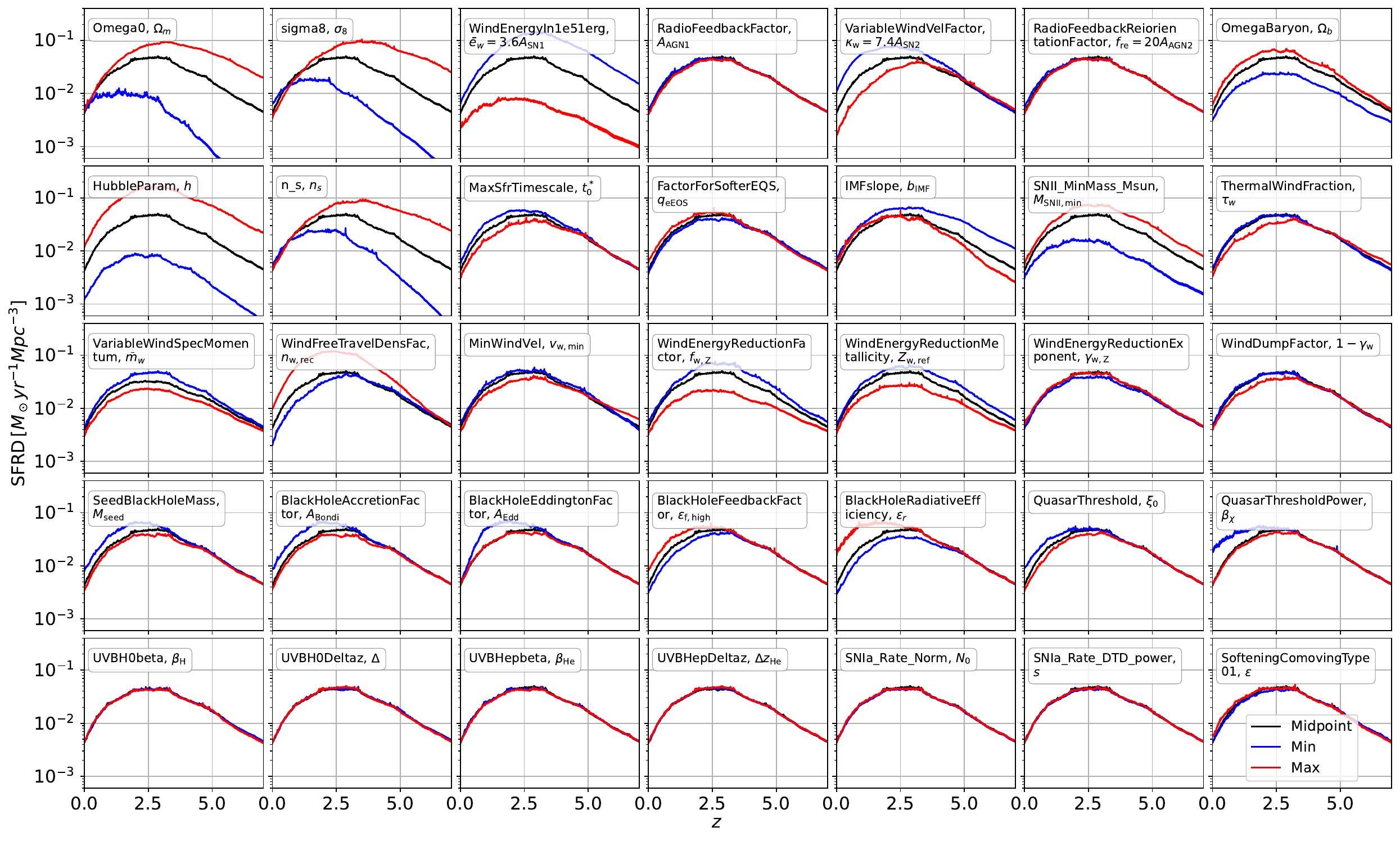}
\caption{Star formation rate density as a function of redshift in the $50\hMpc$ 1P simulation set, showing the effect of varying individual parameters, similarly to \autoref{fig:1P_Pk}. Changes in cosmological and stellar feedback parameters in particular modulate the amplitude and peak of the cosmic star formation history.}
\label{fig:1P_SFRD}
\end{figure*}

\begin{figure*}[ht!]
\includegraphics[width=0.99\textwidth]{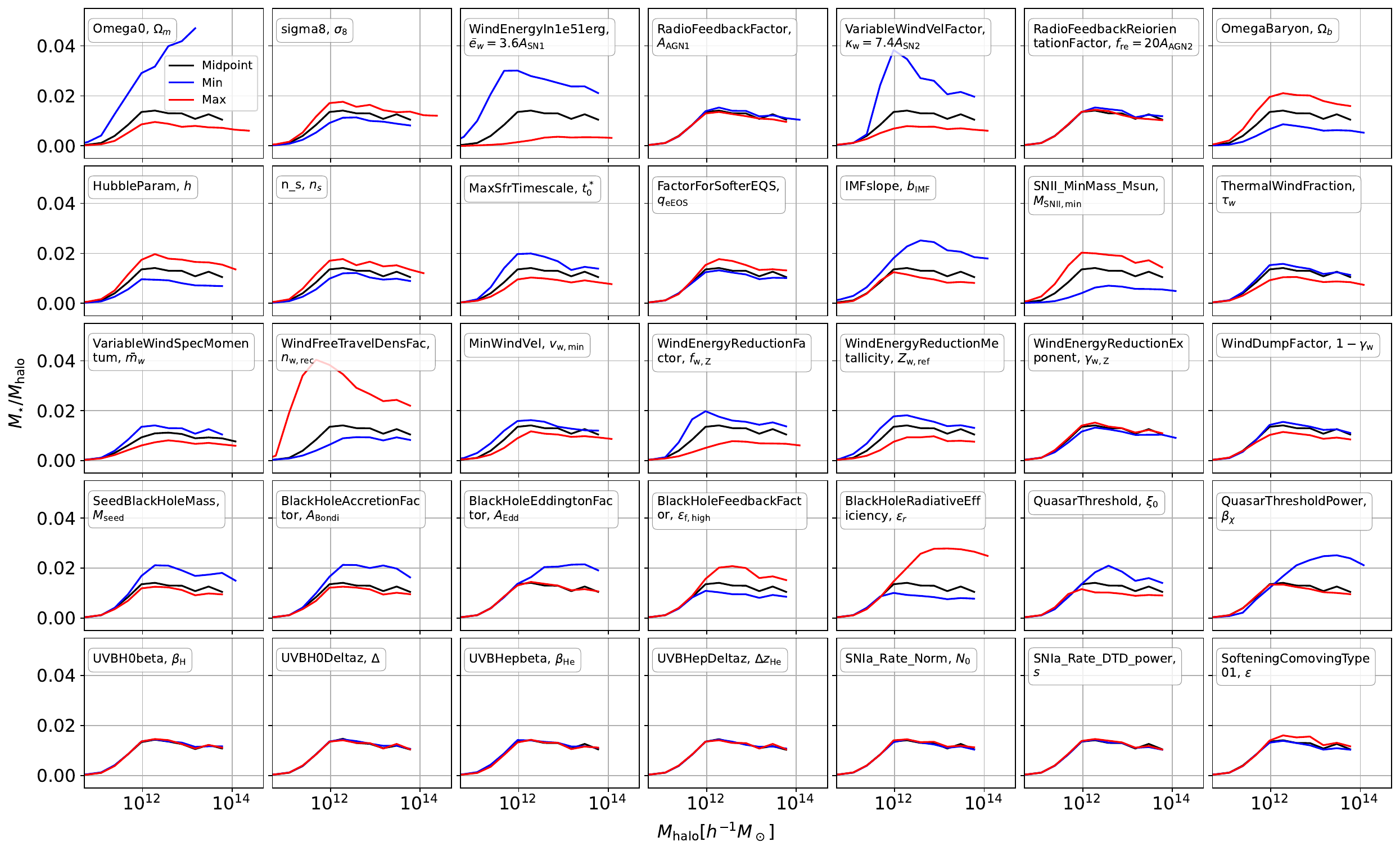}
\caption{Stellar-to-halo mass relation at $z=0$ for the $50\hMpc$ 1P simulation set, illustrating how variations in cosmological and astrophysical parameters affect galaxy formation efficiency.}
\label{fig:1P_MsMh}
\end{figure*}

In this section, we present a selection of the quantities shown in Figures \ref{fig:global_properties} and \ref{fig:galaxy_scaling_relations}, or close variants thereof, in terms of their dependencies on each of the 35 parameters that are varied in our new simulations. We do so utilizing the 1P set, which includes 5 variations for each parameter, including the common fiducial setup. For most parameters, the fiducial parameter value is in the middle (in either linear or log space) between the minimum and maximum values, but for a few parameters (see \autoref{tab:parameters}), the fiducial value is intermediate but not in the middle, or is even equal to the value at one of the extremes. In Figures \ref{fig:1P_Pk} through \ref{fig:1P_MsMh}, we show the results from three values for each parameter: the minimum (blue), the maximum (red), and the middle point (black), regardless of the fiducial value -- though for most parameters, the latter is also the fiducial.

\autoref{fig:1P_Pk} presents the ratio of the matter power spectra between the hydrodynamical simulations and their corresponding N-body simulations. Almost everywhere in the parameter space, the power is suppressed on scales $k\approx1-50h/\Mpc$ due to the presence of baryonic physics and feedback, and then turns into enhancement at smaller, galactic scales. The cosmological parameters $\Omega_m$, $\Omega_b$, and $n_s$ induce large variations, while the leading astrophysical parameters with the largest effect include several of the AGN feedback parameters, such as $\epsilon_r$ and $\beta_\chi$, which, at their respective maximum and minimum values, lead to a strong suppression of kinetic-mode AGN feedback that, in turn, causes a power enhancement instead of suppression at $k\gtrsim1h/\Mpc$. The stellar physics-related parameters $\bar{e}_w$ and $b_{\rm IMF}$ also have a substantial effect, operating through their modulation of black hole masses and, hence, of AGN feedback \citep{LeeM_24a,LinS_25a}. A comparison of the scale of these variations to the average CV spread of $\approx0.03$ (\autoref{fig:CV}) makes it clear that the power spectrum ratio has no constraining power on many of the parameters, but barring degeneracies, it has the potential to constrain several others.

Some parameters affect the power spectrum suppression by shifting it up and down in amplitude, while others affect it by shifting it up and down in wavenumber, i.e.~horizontally in \autoref{fig:1P_Pk}. Additionally, some parameters have more idiosyncratic effects. Most parameters elicit asymmetric responses around the middle value, and some parameter effects appear to be close to saturation in the middle value, i.e.~they elicit responses primarily in one direction of the variations. All in all, the responses appear to be largely unique to each parameter; namely, if there are degeneracies between different parameters regarding their effects on the power spectrum suppression, they are more complex than those between merely pairs of individual parameters.

The cosmic star-formation history, shown in \autoref{fig:1P_SFRD}, is barely sensitive to AGN feedback at $z\gtrsim3$, and even at later times, the sensitivity to BH parameters is modest because massive halos have a limited contribution to star-formation even in the absence of AGN feedback. However, in addition to a superlinear sensitivity to the cosmological parameters that control the number of halos in the box (all except $\Omega_b$), this quantity displays significant sensitivities to several galactic wind parameters such as $\bar{e}_w$, $\kappa_{\rm w}$, and $M_{\rm SNII,min}$. Similarly to the case of the power spectrum suppression, few parameter pairs appear to be fully degenerate, with individual parameters largely having unique signatures on the cosmic star-formation history. A comparison of the scale of these variations to the average CV spread of $\approx0.04\dex$ (\autoref{fig:CV}) suggests that the cosmic star-formation history should have more constraining power for astrophysical parameters at later cosmic times, since that is when the cosmic variance is smaller while the parameter dependencies are larger. Its dependencies on the cosmological parameters, in contrast, tend to be larger at earlier times, leaving the door open for constraining power despite larger cosmic variance.

\autoref{fig:1P_MsMh} shows that almost all of the 28 parameters that were varied in SB28 (top four rows) have substantial effects on the stellar-to-halo mass ratio, including cosmological, stellar, and black hole parameters. The black hole parameters, in particular those shown in the fourth row, tend to affect mostly halos with masses above $10^{12}\hMsun$, as expected, while other parameters have a wider range of influence. A comparison of the scale of these variations to the average CV spread of $\approx0.0005$ (\autoref{fig:CV}) suggests that most of these 28 parameters are likely to be constrained (up to degeneracies) by this quantity.

In contrast, the 7 new parameters that are varied in the new SB35 set (bottom row) have marginal effects not only on the stellar-to-halo mass ratio but also on the quantities shown in Figures \ref{fig:1P_Pk} and \ref{fig:1P_SFRD}. While this set of three quantities is certainly not exhaustive, its diversity suggests that these additional 7 parameters, including the gravitational softening $\epsilon$, which is the first numerical (as opposed to subgrid physics) parameter varied in CAMELS, have limited effects overall. We do see (but do not show explicitly here) that the new $N_0$ and $s$, which control SNIa rates, do indeed give rise to changes in [Mg/Fe] ratios, and that (as shown in \autoref{fig:T0_gamma}) the radiation background parameters affect properties of the IGM. These limited effects suggest that the expansion of the parameter space with respect to SB28 is likely not the driving factor in the less-than-na\"ive scaling of the inference results presented in \autoref{sec:inference} with the volume difference between the SB35 and SB28 sets.

\setbox0=\hbox{\includegraphics{L250999.png}}
\setbox0=\hbox{\includegraphics{L500999.png}}
\setbox0=\hbox{\includegraphics{L50_L25_Nhalos_above_1e13Msunhinv_python_v3.png}}
\setbox0=\hbox{\includegraphics{mass_distribution_results_10_v5.pdf}}

\end{document}